\documentclass[a4paper]{article}
\usepackage{amsmath,amssymb,theorem,fullpage}
\usepackage{hyperref}
\usepackage[nosort]{cite}

\let\frac\undefined

\allowdisplaybreaks

\numberwithin{equation}{section}

\def\Maketitle{{\def\newpage{}\maketitle}}

\def\eq#1{\begin{equation}#1\end{equation}}
\long\def\subeq#1{\begin{subequations}#1\end{subequations}}

\def\Align#1{\begin{align}#1\end{align}}

\def\AlignAt#1#2{\begin{alignat}{#1}#2\end{alignat}}
\def\Aligned#1{\begin{aligned}#1\end{aligned}}
\def\Gather#1{\begin{gather}#1\end{gather}}
\def\Gathered#1{\begin{gathered}#1\end{gathered}}
\def\Multline#1{\begin{multline}#1\end{multline}}
\def\Matrix#1{\begin{matrix}#1\end{matrix}}

\def\Cases#1{\begin{cases}#1\end{cases}}
\def\?{\notag}
\def\d{\partial}

\def\Re{\mathop{\rm Re}\nolimits}
\def\Im{\mathop{\rm Im}\nolimits}
\def\sign{\mathop{\rm sign}\nolimits}
\def\Res{\mathop{\rm Res}}
\def\const{\mathop{\rm const}\nolimits}
\def\cA{{\cal A}}
\def\cB{{\cal B}}
\def\cC{{\cal C}}
\def\cD{{\cal D}}

\def\cG{{\cal G}}

\def\cI{{\cal I}}
\def\cL{{\cal L}}
\def\cO{{\cal O}}

\def\ve{\varepsilon}
\def\sh{\mathop{\rm sh}\nolimits}
\def\ch{\mathop{\rm ch}\nolimits}

\def\tg{\mathop{\rm tg}\nolimits}

\def\C{{\mathbb{C}}}

\def\Z{{\mathbb{Z}}}

\def\e{{\rm e}}
\def\i{{\rm i}}

\def\tV{{\tilde V}}

\def\bc{{\bar c}}
\def\bh{{\bar h}}
\def\bcA{{\bar{\cal A}}}
\def\hmu{{\hat\mu}}

\def\lvac{\langle{\rm vac}}
\def\rvac{{\rm vac}\rangle}
\def\Ress#1{\mathop{\Res\hphantom{_{_{#1}}}}_{a=a_{#1}}}

\makeatletter

\def\section{\@startsection{section}{1}{\z@}%
                                   {-3.5ex \@plus -1ex \@minus -.2ex}%
                                   {2.3ex \@plus.2ex}%
                                   {\normalfont\normalsize\bfseries}}
\def\subsection{\@startsection{subsection}{2}{\z@}%
                                     {-3.25ex\@plus -1ex \@minus -.2ex}%
                                     {1.5ex \@plus .2ex}%
                                     {\normalfont\normalsize\bfseries\itshape}}
\def\@seccntformat#1{\csname the#1\endcsname.~~}
\long\def\@makecaption#1#2{%
  \vskip\abovecaptionskip
  \sbox\@tempboxa{\small#1. #2}%
  \ifdim \wd\@tempboxa >0.9\hsize
  {\leftskip=0.05\hsize\rightskip=0.05\hsize\relax\small
    #1. #2\par}
  \else
    \global \@minipagefalse
    \hb@xt@\hsize{\hfil\box\@tempboxa\hfil}%
  \fi
  \vskip\belowcaptionskip}
\def\Appendix{\appendix
  \def\@seccntformat##1{Appendix~\csname the##1\endcsname.~~}}

\let\over\@@over
\let\atop\@@atop
\let\above\@@above
\let\overwithdelims\@@overwithdelims
\let\atopwithdelims\@@atopwithdelims
\let\abovewithdelims\@@abovewithdelims

\makeatother

\begin{document}


\title{Resonances in sinh- and sine-Gordon models\\
and higher equations of motion in Liouville theory}

\author{Michael Lashkevich\\[\medskipamount]
\parbox[t]{0.9\textwidth}{\normalsize\it\raggedright
Landau Institute for Theoretical Physics,
142432 Chernogolovka of Moscow Region, Russia}}

\date{}

\Maketitle

\begin{abstract}
The notion of operator resonances was introduced earlier by Al.~Zamolodchikov within the framework of the conformal perturbation theory. The resonances are related to logarithmic divergences of integrals in the perturbation expansion, and manifest themselves in poles of the correlation functions and form factors of local operators considered as functions of conformal dimensions. The residues of the poles can be computed by means of some operator identities. Here we study the resonances in the Liouville, sinh- and sine-Gordon models, considered as perturbations of a massless free boson. We show that the well-known higher equations of motion discovered by Al.~Zamolodchikov in the Liouville field theory are nothing but resonance identities for some descendant operators. The resonance expansion in the vicinity of a resonance point provides a regularized version of the corresponding operators. We try to construct the corresponding resonance identities and resonance expansions in the sinh- and sine-Gordon theories. In some cases it can be done explicitly, but in most cases we are only able to obtain a general form so far. We show nevertheless that the resonances are perturbatively exact, which means that each of them only appears in a single term of the perturbation theory.
\end{abstract}

\section{Introduction}
\label{Sec:Introduction}

The notion of resonances in the conformal perturbation theory was introduced by Al.~Zamolodchikov in~\cite{Zamolodchikov:1990bk}. He considered a conformal field theory perturbed by a relevant perturbation, i.~e.\ a perturbation described by the action $S_{\rm pert}\sim\int d^2x\,\Phi_{\rm pert}(x)$, where $\Phi_{\rm pert}(x)$ is a local spinless field of conformal dimension $\Delta_{\rm pert}<1$. Let $\cO(x)$ be any field of given conformal dimension $\Delta$. Suppose that the fusion rule of the $\cO$ operator and $k$ instances of the $\Phi_{\rm pert}$ operators contain an operator $\cO'(x)$ of the dimension $\Delta'$. Then, if $\Delta\ge\Delta^{\rm res}\equiv\Delta'+k(1-\Delta_{\rm pert})$, ultraviolet divergences take place in the $k$th order of the perturbation theory. To avoid the divergence, the operator $\cO$ must be renormalized by adding the operator $\cO'$ with an infinite coefficient. Alternatively, the renormalization can be often achieved by analytic continuation from the region of convergence. The most subtle situation takes place on the threshold of the ultraviolet divergences $\Delta=\Delta^{\rm res}$. In this case the divergence is logarithmic and the definition of the renormalized field $\cO^{\rm ren}$ is ambiguous. Another manifestation of this phenomenon is that every correlation function of the operator $\cO$ possesses a pole at the point $\Delta=\Delta^{\rm res}$. The residue of the pole coincides with a correlation function of the operator $\cO'$ so that one can write the identity as an operator one:
\eq{
\Res_{\Delta=\Delta^{\rm res}}\cO(x)=\const\times\cO'(x).
\label{Intr:res}
}
This phenomenon plays an important role in the two-dimensional integrable quantum field theory. Exact vacuum expectation values and form factors are given by analytic functions in the conformal dimensions of the corresponding operators. These functions possess poles at the resonance points~\cite{Fateev:1998xb,Fateev:2009kp}. In order to extract some physically significant information at these points, the singularities must be cancelled by an appropriate renormalization procedure.

Another famous result by Al.~Zamolodchikov is the discovery of the higher equations of motion in the Liouville field theory~\cite{Zamolodchikov:2003yb}. The Liouville theory is a theory of a scalar real boson field $\varphi(x)$ with the potential of the form $\e^{b\varphi}$. Being a conformal field theory it contains two Virasoro algebras generated by the modes $\cL_k$ and $\bar\cL_k$ of the two nonzero components of the energy momentum tensor $T_{--}$ and $T_{++}$. The higher equations of motion are operator relations of the form
$$
D_{mn}\bar D_{mn}(\varphi\,\e^{a_{mn}\varphi})=2B_{mn}\e^{a_{m,-n}\varphi}.
$$
Here $a_{mn}={1-m\over2}b^{-1}+{1-n\over2}b$, $B_{mn}$ are some constants, and $D_{mn}$ and $\bar D_{mn}$ are polynomials in the modes $\cL_{-k}$ and $\bar\cL_{-k}$ ($k>0$) correspondingly of the level $mn$, such that $D_{mn}\e^{a_{mn}\varphi}=\bar D_{mn}\e^{a_{mn}\varphi}=0$. The simplest case $m=n=1$ reduces to the usual equation of motion. In the present paper the higher equations of motion are interpreted as resonance identities of the form (\ref{Intr:res}) in the perturbation theory from the free massless boson with the perturbation $\int d^2x\,\e^{b\varphi}$, and the corresponding renormalized operators are defined.

The sinh-Gordon theory can be considered either as a relevant perturbation of the Liouville theory or as a relevant perturbation of the free massless boson. One may expect that it inherits the resonances of the Liouville theory, and, besides, possesses its own resonances. We shall show how the resonance identities are modified in the sinh-Gordon theory. It turns out that if either $m$ or $n$ is even the resulting resonance identity coincides with that in the Liouville theory, while in the case of odd $m$ and $n$ additional terms appear. These additional terms can be easily written explicitly, if either $m=1$ or $n=1$. Further, we try to generalize the results to the sine-Gordon theory. The situation in that case becomes more involved due to the fact that the very operator $\cO'$ can possess a resonance pole. We use the form factor analyticity conjecture to understand the structure in this case.

Note that one of the main motivations to this study was caused by the present unsatisfactory state of the form factor theory. Within the bootstrap approach we try to identify solutions to the form factor equations~\cite{Smirnov:1992vz} with some particular operators in the Lagrangian or conformal perturbation theory, and ignore the fact that we do not clearly understand the very sense of these operators beyond the free field and conformal field theory.

In Sec.~\ref{Sec:Liouville&HEM} we recall some basic facts concerning the Liouville field theory and the higher equations of motion. In Sec.~\ref{Sec:HEM&resonnances} we demonstrate the connection of the higher equations of motion with the resonance identities. The generalizations to the sinh- and sine-Gordon models are made in Secs.~\ref{Sec:Res-shG} and~\ref{Sec:Res-sG} correspondingly. In Appendix~\ref{App:log} the logarithmic behavior of some particular fields in the Liouville theory is substantiated. Appendix~\ref{App:FF-descendants} contains some support of the results of the paper by the form factor theory.

\section{Liouville field theory and higher equations of motion}
\label{Sec:Liouville&HEM}

Here we recall some basic facts concerning the Liouville field theory and, in particular, what is called the higher equations of motion. The Liouville field theory on the Minkowski plane is defined by the action
\eq{
S_{\rm L}[\varphi]=\int d^2x\,\left({(\d_\mu\varphi)^2\over16\pi}-\mu\,\e^{b\varphi}\right).
\label{Liouville-action}
}
We shall also use the letters $Q$ and $p$ defined as
\eq{
Q=b+b^{-1},
\qquad
p=Q^{-1}b,
\qquad
b^2={p\over1-p}.
\label{Qpdef}
}
Besides, we shall always use the light-cone coordinates and derivatives:
$$
x^\mp=x^1\mp x^0,
\qquad
\d_\mp={\d\over\d x^\mp}.
$$
The space of local operators of the Liouville theory is spanned by the exponential operators
\eq{
V_a(x)=\e^{a\varphi(x)}
\label{Vadef}
}
and their descendants
\eq{
\d_-^{k_1}\varphi\cdots\d_-^{k_m}\varphi\,\d_+^{l_1}\varphi\cdots\d_+^{l_n}\varphi\,V_a(x).
\label{Va-desc-def}
}
The pair of numbers $(K,L)=\left(\sum k_i,\sum l_i\right)$ is called the level of a descendant. The numbers $K$ and $L$ are called its chiral (right and left correspondingly) levels. The descendants with $L=0$ are called right, while those with $K=0$ left. The (Lorentz) spin of the operator is equal to $K-L$.

It is known that not all the operators (\ref{Va-desc-def}), and even not all the exponential operators (\ref{Vadef}), are different. The exponential operators are pairwise identified due to the reflection relation~\cite{Zamolodchikov:1995aa}
\eq{
V_a(x)=R_aV_{Q-a}(x).
\label{RR}
}
The function $R_a$ is called the reflection function and given by
\eq{
R_a=-M^{2Q(Q-2a)}
{\Gamma(2ba-b^2)\Gamma(2b^{-1}a-b^{-2})\over\Gamma(2b(Q-a)-b^2)\Gamma(2b^{-1}(Q-a)-b^{-2})},
\label{Radef}
}
where
\eq{
M=\hmu^{1\over2(1+b^2)},
\qquad
\hmu={\pi\Gamma(b^2)\over\Gamma(1-b^2)}\mu.
\label{Mdef}
}
This identification imposes an identification of the descendant operators, which will be described a little later.

Since these basic elements depend on the continuous parameter $a$, a complete description of the space of local operators must admit linear combinations involving integrating and derivatives in~$a$. In particular, we may consider the operators
$$
V'_a(x)={dV_a(x)\over da}
$$
and their descendants $\d_-^{k_1}\varphi\cdots\d_-^{k_m}\varphi\,\d_+^{l_1}\varphi\cdots\d_+^{l_n}\varphi\,V'_a(x)$.

The Liouville field theory possesses the conformal symmetry, which is preserved on the quantum level. It means the stress-energy tensor $T^\mu_\nu$ can be rendered traceless by adding some full derivatives in coordinates. Hence, in the light-cone coordinates we have $T_{+-}=T_{-+}=0$. For two other components we shall use the standard notation
\eq{
\left.\Matrix{T(x)\\\bar T(x)}\right\}=-2\pi T_{\mp\mp}(x)=-{1\over4}(\d_\mp\varphi)^2+{Q\over2}\d_\mp^2\varphi.
\label{Tmpdef}
}
Due to the energy-momentum conservation the component $T$ is a function of $x^-$ only, being $x^+$-\hskip0pt independent, while the component $\bar T$ is a function of~$x^+$. This defines two Virasoro algebras as follows. Consider the Wick rotation $x^0=-\i\tau$. The light-cone variables becomes complex coordinates on the Euclidean plane: $x^-=z=x^1+\i\tau$, $x^+=\bar z=x^1-\i\tau$. One can define the operators $\cL_n$, $\bar\cL_n$ acting on local operators~$\cO(x)$:
\eq{
\cL_n\cO(x)=\oint{dw\over2\pi\i}\,(w-z)^{n+1}T(w)\cO(x),
\qquad
\bar\cL_n\cO(x)=\oint{d\bar w\over2\pi\i}(\bar w-\bar z)^{n+1}\bar T(\bar w)\cO(x).
\label{cLdef}
}
These operators form the Virasoro algebra
$$
[\cL_m,\cL_n]=(m-n)\cL_{m+n}+{c\over12}m(m^2-1)\delta_{m+n,0},
$$
and the same algebra for $\bar\cL_n$, while $[\cL_m,\bar\cL_n]=0$. Both right and left algebras possess the same central charge
$$
c=1+6Q^2.
$$

With respect to the Virasoro algebras the exponential operators $V_a$ are primary,
$$
\cL_nV_a=\bar\cL_nV_a=0\quad(n>0),
\qquad
\cL_0V_a=\bar\cL_0V_a=\Delta_aV_a,
$$
with the conformal dimensions
$$
\Delta_a=a(Q-a).
$$
The descendants (\ref{Va-desc-def}) for generic values of $a$ can be expressed as linear combinations of the conformal basis:
\eq{
\cL_{-k_1}\cdots\cL_{-k_m}\bar\cL_{-l_1}\cdots\bar\cL_{-l_n}V_a(x),
\label{Va-desc-conf}
}
where the level $(K,L)$ coincides with $\left(\sum k_i,\sum l_i\right)$. Note that the reflection property for the descendant operators has a simple form in this basis. Namely,
\eq{
\cL_{-k_1}\cdots\cL_{-k_m}\bar\cL_{-l_1}\cdots\bar\cL_{-l_n}V_a(x)
=R_a\cL_{-k_1}\cdots\cL_{-k_m}\bar\cL_{-l_1}\cdots\bar\cL_{-l_n}V_{Q-a}(x).
\label{RR-desc}
}

The particular operators $\cL_{-1}$ and $\bar\cL_{-1}$ always coincide with the light-cone derivatives:
$$
\cL_{-1}=\d_-,
\qquad
\bar\cL_{-1}=\d_+.
$$
Thus, the standard equation of motion
\eq{
\d_-\d_+\varphi=2\pi\mu b\e^{b\varphi}
\label{Liouville-EM}
}
can be rewritten as
\eq{
\cL_{-1}\bar\cL_{-1}V'_0(x)=2\pi\mu b\,V_b(x).
\label{Liouville-eqmot-cL}
}
In the quantum theory the equation of motion does not determine the behavior of the model as a whole. It just establishes a relation between several operators. Al.~Zamolodchikov proposed~\cite{Zamolodchikov:2003yb} a generalization of the equation of motion in the Liouville field theory. Let us recall his main result.

Consider the special type of exponential operators
\eq{
V_{mn}(x)=V_{a_{mn}}(x),
\qquad
a_{mn}={1-m\over2}b^{-1}+{1-n\over2}b,
\qquad
m,n\in\Z.
\label{amndef}
}
For $a=a_{mn}$ and $m,n>0$, the conformal dimensions coincide with the Kac dimensions~\cite{Kac:1979:LNP} and the module of the Virasoro algebra generated by $\cL_{-k}$, $\bar\cL_{-k}$ ($k>0$) on the operator $V_{mn}(x)$ as on the highest weight vector is non-free on the level~$mn$. It means that there exist nonzero operators of the form
\eq{
\Aligned{
D_{mn}
&=\sum^{mn}_{s=1}\sum_{1\le k_1\le\cdots\le k_s\atop k_1+\cdots+k_s=mn}
A_{k_1\ldots k_s}\cL_{-k_1}\cdots\cL_{-k_s},
\\
\bar D_{mn}
&=\sum^{mn}_{s=1}\sum_{1\le k_1\le\cdots\le k_s\atop k_1+\cdots+k_s=mn}
A_{k_1\ldots k_s}\bar\cL_{-k_1}\cdots\bar\cL_{-k_s}
}\label{Dmndef}
}
with some real coefficients $A_{\dots}$, such that
\eq{
D_{mn}V_{mn}=\bar D_{mn}V_{mn}=0,
\qquad
m,n>0.
\label{DV=0}
}
It is convenient to normalize these operators by the condition $A_{1\ldots1}=1$, so that
$$
D_{mn}=\cL_{-1}^{mn}+\cdots,
\qquad
\bar D_{mn}=\bar\cL_{-1}^{mn}+\cdots.
$$
The higher equations of motion read
\eq{
D_{mn}\bar D_{mn}V'_{mn}(x)=2B_{mn}V_{m,-n}(x).
\label{HEM}
}
Here the coefficients $B_{mn}$ are explicitly known
\eq{
B_{mn}=\hmu^n\cB_{mn}
=\hmu^nb^{1+2n-2m}{\Gamma(n-mb^2)\over\Gamma(1-n+mb^2)}
\prod_{1-m\le k\le m-1,\ 1-n\le l\le n-1\atop(k,l)\ne(0,0)}(kb^{-1}+lb).
\label{Bmn}
}

\section{Higher equations of motion as resonance identities}
\label{Sec:HEM&resonnances}

Let us recall the phenomenon of resonances between operators in the conformal perturbation theory~\cite{Zamolodchikov:1990bk,Fateev:1998xb,Fateev:2009kp}. Consider a field theory in two dimensions that can be represented as a perturbation of a conformal field theory by a relevant primary operator:
\eq{
S=S_0+S_{\rm pert}=S_0-g\int d^2x\,\Phi_{\rm pert}(x),
\label{genaction}
}
where $S_0$ is the action of a conformal field theory and $\Phi_{\rm pert}(x)$ is a spin 0 primary conformal operator of conformal dimension $\Delta_{\rm pert}<1$. The coupling constant $g$ has the dimensionality $({\rm mass})^{2(1-\Delta_{\rm pert})}$. Formally, any (chronologically ordered) correlation function $\langle\cdots\rangle$ can be represented in terms of the unperturbed correlation function $\langle\cdots\rangle_0$ as
$$
\langle\cO_1(x_1)\cO_2(x_2)\cdots\rangle
={\langle\e^{\i S_1}\cO_1(x_1)\cO_2(x_2)\cdots\rangle_0\over\langle\e^{\i S_1}\rangle_0}.
$$
Since $S_{\rm pert}$ is proportional to a ``small'' parameter $g$, the r.~h.~s.\ can be expanded in a perturbation series in~$g$. Naturally, an arbitrary term of this expansion consists of terms of the form
$$
\int d^2y_1\cdots d^2y_k\,\langle\Phi_{\rm pert}(y_1)\cdots\Phi_{\rm pert} (y_k)
\cO_1(x_1)\cO_2(x_2)\cdots\rangle_0.
$$
The integrals are calculated by their continuation to the Euclidean spacetime, so that we shall suppose such continuation everywhere, and write $|x|^2=-x^2=x^-x^+$. The integrals may possess both ultraviolet and infrared divergences. We will be interested in the former. Consider the region of the integration over $y_i$ in the vicinity of the point where an operator is inserted, e.~g.~$x=x_1$. Suppose that one of the operators in the correlation function $\cO(x)$ is any, not necessarily primary, spin 0 operator of a given conformal dimension~$\Delta$. For simplicity, consider the case $k=1$. Then applying the operator product expansions we obtain
$$
\int d^2y\,\Phi_{\rm pert}(y)\cO(x)
=\sum_nC_n\int d^2y\,|y-x|^{2(\Delta'_n-\Delta_{\rm pert}-\Delta)}\cO'_n(x),
$$
where $C_n$ are some structure constants and $\Delta'_n$ are conformal dimensions of the operators~$\cO'_n(x)$ that enter the operator product expansion. Operators of nonzero spin, which have different right and left conformal dimensions, do not contribute the expansion in the ultraviolet region. If the dimension $\Delta'_n$ large enough, $\Delta'_n>\Delta+\Delta_{\rm pert}-1$, the corresponding contribution is convergent. If $\Delta'_n<\Delta+\Delta_{\rm pert}-1$, the contribution diverges, but the divergence can be canceled by renormalization of the operator $\cO$ according to $\cO\to\cO^{\rm ren}=\cO+a_ng\Lambda^{2(\Delta_{\rm pert}+\Delta-\Delta'_n-1)}\cO'_n$ with some constant factor~$a_n$ and $\Lambda$ being the ultraviolet cutoff in the momentum space, which will be set to infinity after renormalization. In both cases the renormalized operator can be defined in a unique way. The situation when the resonance condition
$$
\Delta'_n=\Delta+\Delta_{\rm pert}-1
$$
is satisfied turns out to be most difficult. In this case the integral diverges logarithmically and the renormalization
$$
\cO^{\rm ren}(\lambda;x)=\cO(x)+a_ng\left(\log{\Lambda\over\lambda}\right)\cO'_n(x)
$$
inevitably contains some arbitrary but finite parameter~$\lambda$ of the dimensionality of mass. This parameter makes the definition of the operator $\cO^{\rm ren}$ intrinsically ambiguous. The last admits a redefinition
$$
\cO^{\rm ren}(\lambda';x)
=\cO^{\rm ren}(\lambda;x)-a_ng\left(\log{\lambda'\over\lambda}\right)\cO'_n(x).
$$
In terms of correlation functions the logarithmic divergence reveals itself as poles of correlation functions with respect to $\Delta$, while the residues of the poles are proportional to the correlations functions, where the operator $\cO$ is substituted by~$\cO'_n$.

Generally, consider an operator $\cO(x)$ of the dimension $\Delta$ and an operator $\cO'(x)$ of the dimension $\Delta'$ that appears in the expansion in the powers of $y_i-x$ of the operator product
$$
\cO(x)\Phi_{\rm pert}(y_1)\cdots\Phi_{\rm pert}(y_k).
$$
We suppose that $\Delta$ is a continuous parameter,%
\footnote{Surely, there are examples where the set of admissible dimensions is discontinuous. In these cases one may need to use some tricks to stir conformal dimensions to see the poles (see e.~g.~\cite{Fateev:2009kp}). Such theories are not the subject of present consideration.}
so that the conformal dimension $\Delta'$ of the admissible operator is a function of~$\Delta$: $\Delta'=\Delta'_k(\Delta)$. Then every correlation function of $\cO(x)$ possesses a pole at the resonance point $\Delta=\Delta_*$, which is a solution to the equation
\eq{
\Delta_*=\Delta^{\rm res}_k(\Delta_*),
\qquad
\Delta^{\rm res}_k(\Delta)\stackrel{\rm def}=\Delta'_k(\Delta)+k(1-\Delta_{\rm pert}).
\label{rescond}
}
The residue of the pole in a correlation function of $\cO$ is proportional to the corresponding correlation function of the operator~$\cO'$, so that we may write it as an operator identity
\eq{
\Res_{\Delta=\Delta_*}\cO(\Delta;x)=ag^k\cO'_*(x).
\label{residue}
}

There are two ways to define the renormalized operator at the residue point. The first way is the `traditional' one by means of cutting off integrations and subtracting the divergent part:
$$
\tilde\cO^{\rm ren}_*(\lambda;x)=\cO_*(x)+2ag^k\left(\log{\Lambda\over\lambda}\right)\cO'_*(x).
$$
The parameter $\lambda$ can be taken, in principle, arbitrary. It means that the logarithmic divergence leads to an ambiguity:
$$
\tilde\cO^{\rm ren}_*(\lambda';x)
=\tilde\cO^{\rm ren}_*(\lambda;x)-2ag^k\left(\log{\lambda'\over \lambda}\right)\cO'_*(x).
$$

The other way uses the expansion in the vicinity of the pole:
\eq{
\cO(x)
={ag^k\lambda^{2(\Delta-\Delta^{\rm res}_k(\Delta))}\cO'(x)\over\Delta-\Delta^{\rm res}_k(\Delta)}
+\cO^{\rm ren}_*(\lambda,x)+O(\Delta-\Delta^{\rm res}_k(\Delta)),
\label{res-expansion}
}
where $\lambda$ is again an arbitrary constant of the dimension of mass. The factor $\lambda^{2(\Delta-\Delta^{\rm res}_k)}$ in the first term is introduced to equate the dimension of the r.~h.~s.\ to that of the l.~h.~s. The arbitrariness of $\lambda$ results in the ambiguity of the definition of the renormalized operator.

Both approaches lead to essentially the same result. Namely, there exists a constant $\kappa$ such that
$$
\cO^{\rm ren}_*(\lambda;x)=\tilde\cO^{\rm ren}_*(\kappa\lambda;x).
$$
The value of $\lambda$ is at our disposal. If the perturbed theory is non-conformal, it possesses a mass scale $M\sim g^{1/2(1-\Delta_{\rm pert})}$, and we have to take $\lambda\propto M$ with an arbitrary dimensionless proportionality constant. In the conformal case we are not bound even by this limitation.

Let us stress that the expansion (\ref{res-expansion}) is taken over the powers of $\Delta-\Delta^{\rm res}_k(\Delta)$ rather than of $\Delta-\Delta_*$, and the numerator of the first term contains the operator $\cO'(x)$, which is a function of the parameter $\Delta$, rather than the resonance operator $\cO'_*(x)$. This form of the expansion is dictated by the structure of the perturbative series~\cite{Zamolodchikov:1990bk} and essentially affects the definition of the renormalized operator, as we shall see in a while.

Now we want to consider the Liouville field theory as a perturbation of the free boson theory:
\eq{
S_0[\varphi]=\int d^2x\,{(\d_\mu\varphi)^2\over16\pi},
\qquad
\Phi_{\rm pert}=\e^{b\varphi},
\qquad
g=\mu.
\label{Liouville-pert}
}
Indeed, in the free boson theory the Virasoro algebras are given by the generators $\cL^0_n$, $\bar\cL^0_n$ obtained from the energy-momentum tensor components
\eq{
\left.\Matrix{T^0\\\bar T^0}\right\}=-2\pi T^0_{\mp\mp}=-{1\over4}(\d_\mp\varphi)^2.
\label{T0mpdef}
}
With respect to this energy-momentum tensor the exponential operators $V_a(x)$ are primary as well, but they possess the conformal dimensions
\eq{
\Delta^0_a=-a^2.
\label{Delta0adef}
}
In particular, the conformal dimension of the perturbation field is
\eq{
\Delta_{\rm pert}=\Delta^0_b=-b^2.
\label{Delta0pert}
}

Consider the operator $D_{mn}\bar D_{mn}V_a(x)$ in the vicinity of the point $a=a_{mn}$. Precisely at this point it vanishes. It is easy to show that in the framework of the free boson field theory
\eq{
D_{mn}\bar D_{mn}V_a(x)=O((a-a_{mn})^2)
\qquad
\text{(free boson).}
\label{DDV-FF}
}
Indeed, the free massless field $\varphi(x)=\varphi(x^-,x^+)$ can be split (up to some zero mode, which does not interest us here) into the sum $\varphi(x^-,x^+)=\varphi_-(x^-)+\varphi_+(x^+)$. Hence, the operator $D_{mn}\bar D_{mn}V_a(x)$ factorizes into the product $(D_{mn}\e^{a\varphi_-(x^-)})(\bar D_{mn}\e^{a\varphi_+(x^+)})$. Each of the two factors is an exponential $\e^{a\varphi_\pm}$ multiplied by a polynomial in the variable~$a$. As far as it vanishes at $a=a_{mn}$ it is of the order $O(a-a_{mn})$ as $a\to a_{mn}$, so that the product of two factors gives rise to~(\ref{DDV-FF}).

Taking into account (\ref{DDV-FF}), it is natural to define the operator
\eq{
\Phi_{a|mn}(x)={D_{mn}\bar D_{mn}V_a(x)\over(a-a_{mn})^2},
\label{Phi-amn-def}
}
which remains finite in the free boson theory in terms of derivatives of the field $\varphi$ at the point $a=a_{mn}$. Its conformal dimension is equal to $\Delta^0_{a|mn}=-a^2+mn$. Its $n$ folded fusion with the perturbation operator $V_b$ generates operators from the Fock module over the exponential operator $V_{a+nb}$ of the conformal dimension $\Delta^0_{a+nb}=-(a+nb)^2$. It is easy to see that the resonance condition
\eq{
\Delta^0_{a|mn}=\Delta^0_{a+nb}+n(1+b^2)
\label{Liouville-rescond}
}
is just satisfied as $a=a_{mn}$. In the vicinity of this point
$$
\Delta-\Delta^{\rm res}_n(\Delta)=2nb(a-a_{mn})
$$
so that we may substitute residues in $\Delta$ by residues in~$a$. We see, that the higher equation of motion (\ref{HEM}) takes the form of a resonance identity:
\eq{
\Ress{mn}\Phi_{a|mn}(x)=2B_{mn}V_{m,-n}(x).
\label{HEM-res}
}
This trivial fact will be the basis for obtaining resonance identities for the sinh- and sine-Gordon models below.

The expansion (\ref{res-expansion}) gives rise to
\eq{
\Phi_{a|mn}(x)
={2B_{mn}\lambda^{4nb(a-a_{mn})}V_{a+bn}(x)\over a-a_{mn}}+\tilde\Phi_{mn}(\lambda;x)+O(a-a_{mn}),
\label{tildePhi-def}
}
providing the definition of the renormalized operator~$\tilde\Phi_{mn}$. A remarkable feature of the Liouville theory is that there is a nice way to get rid of the ambiguity. We can benefit from the transformation
\eq{
\varphi\to\varphi+\xi,
\qquad
\mu\to\mu\e^{-b\xi}.
\label{xi-trans}
}
Suppose the parameter $\lambda$ to be proportional to $\mu^\delta$. We know that under the transformation (\ref{xi-trans}) the l.~h.~s.\ transforms as
$$
\Phi_{a|mn}\to\Phi_{a|mn}\e^{a\xi},
$$
while the first term in the r.~h.~s.\ as
$$
B_{mn}\lambda^{4nb(a-a_{mn})}V_{a+nb}
\to B_{mn}\lambda^{4nb(a-a_{mn})}V_{a+nb}\e^{(a-4\delta nb^2(a-a_{mn}))\xi}.
$$
Comparing both sides of (\ref{tildePhi-def}) we get the transformation rule
$$
\tilde\Phi_{mn}(\kappa\mu^\delta;x)
\to(\tilde\Phi_{mn}(\kappa\mu^\delta\e^{-\delta b\xi};x)+8\xi\delta nb^2B_{mn}V_{m,-n}(x))\e^{a_{mn}\xi}.
$$
The simple transformation rule $\tilde\Phi_{mn}\to\tilde\Phi_{mn}\e^{a_{mn}\xi}$ only restores in the case $\delta=0$. Thus, $\lambda$ is dimensionless and may be set to~1. The operator
\eq{
\Psi_{mn}(x)=\tilde\Phi_{mn}(1;x)
\label{Psidef-Liouville}
}
with the transformation rule
$$
\Psi_{mn}(x)\to\Psi_{mn}(x)\e^{a_{mn}\xi}
$$
will be considered as the uniquely defined renormalized form of~$\Phi_{a_{mn}|mn}$.

Now, let us make sure that this definition is consistent with the common sense. For this purpose explicitly consider the simplest case $m=n=1$. In the free boson field theory the field $\Phi_{a|11}$ is given by
\eq{
\Phi_{a|11}=\d_-\varphi\,\d_+\varphi\,\e^{a\varphi}
\qquad
\text{(free boson),}
\label{Phi-a11-boson}
}
since $\d_-\d_+\varphi=0$. Hence, its value at the point $a=a_{11}=0$ is well defined: $\Phi_{0|11}=\d_-\varphi\d_+\varphi$. Our aim is to define the corresponding operator in the Liouville field theory.

In the Liouville field theory the expression is not so simple:
\eq{
\Phi_{a|11}=a^{-2}\d_-\d_+\e^{a\varphi}=(\d_-\varphi\,\d_+\varphi+a^{-1}\d_-\d_+\varphi)\e^{a\varphi}
=a^{-1}\d_-\d_+\varphi\,e^{a\varphi}+\d_-\varphi\,\d_+\varphi+O(a).
\label{Phi-a11-Liouville}
}
According (\ref{tildePhi-def}), (\ref{Psidef-Liouville}) we have
\eq{
\Phi_{a|11}(x)=a^{-1}\cdot2\pi\mu b\e^{(a+b)\varphi(x)}+\Psi_{11}(x)+O(a).
\label{Psi11-def}
}
It can be elicited from the perturbation theory that for small values of~$a$
$$
\d_-\d_+\varphi\,\e^{a\varphi}=2\pi\mu b\e^{(a+b)\varphi}+O(a^2).
$$
Hence, by comparing (\ref{Psi11-def}) with (\ref{Phi-a11-Liouville}) we immediately obtain
\eq{
\Psi_{11}(x)=\d_-\varphi\,\d_+\varphi(x).
\label{dphidphi}
}
This is the point, where the appearance of $\cO'$ instead of $\cO'_*$ in the expansion (\ref{res-expansion}) is crucial. If we put there $\cO'_*$, the renormalized operator would have the form $\d_-\varphi\,\d_+\varphi+\varphi\,\d_-\d_+\varphi$. This could hardly be considered as a renormalized form of the operator (\ref{Phi-a11-boson}) at $a=0$, since the second term is not invariant under the transformation (\ref{xi-trans}) and cannot be cancelled by adding~$\e^{b\varphi}$. Besides, the operator could be extracted from the $a$-expansion of $\d_+\d_-\e^{a\varphi}$ and, hence, it would mean an unnatural reduction of the space of local operators after the perturbation.

Now we want to rederive (\ref{dphidphi}) in a more conventional way. For this purpose let us slightly move apart the points where the $\varphi$ fields are placed in the product $\d_-\varphi\,\d_+\varphi(x)$. Define the renormalized product by subtracting the pair correlation function:
$$
\d_-\varphi\,\d_+\varphi(x)
=(\d_-\varphi(x')\d_+\varphi(x)-\langle\d_-\varphi(x')\d_+\varphi(x)\rangle)|_{x'\to x}.
$$
Let us rewrite it as
\Align{
\d_-\varphi\,\d_+\varphi(x)
&=(\d_-\varphi(x')\d_+\varphi(x)+\langle\varphi(x')\d_-\d_+\varphi(x)\rangle)|_{x'\to x}
\?\\
&=(\d_-\varphi(x')\d_+\varphi(x)+\varphi(x')\d_-\d_+\varphi(x))|_{x'\to x}
-(\varphi(x')\d_-\d_+\varphi(x)-\langle\varphi(x')\d_-\d_+\varphi(x)\rangle)|_{x'\to x}
\label{dphidphi-sum}
}
The first parenthesis in the last line is the second order of the $a$ expansion of $\d_-\d_+\e^{a\varphi}$. It can be rewritten as
\Multline{
(\d_-\varphi(x')\d_+\varphi(x)+\varphi(x')\d_-\d_+\varphi(x))|_{x'\to x}
=a^{-2}\d_-\d_+\left.(\e^{a\varphi}-a\varphi)\right|_{a\to0}
\\
=\left.\left(a^{-2}\d_-\d_+\e^{a\varphi}-a^{-1}\cdot2\pi\mu b\e^{b\varphi}\right)\right|_{a\to0}
=\left.\left(\Phi_{a|11}(x)-a^{-1}\cdot2\pi\mu b\e^{b\varphi(x)}\right)\right|_{a\to0}.
\?}
We used the equation of motion and the definition~(\ref{Phi-a11-Liouville}) of~$\Phi_{a|11}$. In the second parenthesis of (\ref{dphidphi-sum}) we can apply the equation of motion:
\Multline{
(\varphi(x')\d_-\d_+\varphi(x)-\langle\varphi(x')\d_-\d_+\varphi(x)\rangle)|_{x'\to x}
=\left.2\pi\mu b(\varphi(x')\e^{b\varphi(x)}-\langle\varphi(x')\e^{b\varphi(x)}\rangle)\right|_{x'\to x}
\\
=2\pi\mu b\,\varphi\e^{b\varphi}
=\left.2\pi\mu b{\textstyle{d\over da}}\e^{(b+a)\varphi}\right|_{a\to0}.
\?}
Here the second and third equalities are two reasonable definitions of the operator $\varphi\e^{b\varphi}$: as a limit of a separated product and as the $a$ derivative of the exponent. We assume these two definitions to be equivalent.

After subtracting two last expressions we obtain
\Align{
\d_-\varphi\,\d_+\varphi(x)
&=\left.\left(\Phi_{a|11}(x)-a^{-1}\cdot2\pi\mu b\e^{b\varphi}
-2\pi\mu b{\textstyle{d\over da}}\e^{(b+a)\varphi}\right)\right|_{a\to0}
\?\\
&=\left.\left(\Phi_{a|11}(x)-a^{-1}\cdot2\pi\mu b\e^{(b+a)\varphi(x)}\right)\right|_{a\to0}
=\Psi_{11}(x).
\?}
Though the manipulations are not quite rigorous and rely on some assumptions, the answer pays for it. We see that the construction in terms of resonance expansion conforms the common sense by providing reasonable regularizations of the operators defined by the perturbation from the free boson field theory.

Finally, we succeeded to get rid of the ambiguities in renormalized operators in the Liouville field theory. More precisely, we swept the ambiguity under the carpet. It became possible due to the fact that the Liouville theory is a conformal field theory itself and we succeeded in non-dimensionalizing the constant~$\lambda$. We shall see below that in the case of massive field theories the analogs of the operators $\Psi_{mn}$ are intrinsically ambiguous. In particular, their vacuum expectation values contain logarithms of the dimensional coupling constant, so that a dimensional regularizer like the parameter $\lambda$ defined above becomes necessary.

\section{Resonances in the sinh-Gordon model}
\label{Sec:Res-shG}

The sinh-Gordon theory is the field theory with the action
\eq{
S_{\rm shG}[\varphi]=\int d^2x\,\left({(\d_\mu\varphi)^2\over16\pi}-2\mu\,\ch b\varphi\right).
\label{shG-action}
}
It can be considered either as a perturbation of the free boson field theory by the operator $\ch b\varphi$ or a perturbation of the Liouville field theory by the operator $\e^{-b\varphi}$. We shall need both points of view, but by default we shall use the former. Each time we use the second point of view, we shall declare it explicitly.

Since there is no invariance like (\ref{xi-trans}) in the sinh-Gordon theory, the theory is non-conformal and the parameter $\mu$ is a genuine dimensional parameter of the dimensionality $({\rm mass})^{2+2b^2}$. In fact, the theory is massive, so that the parameter $M$ defined in (\ref{Mdef}) is proportional to the mass $m$ of the particle~\cite{Zamolodchikov:1995xk}:
\eq{
m={8\sqrt\pi\,Qb^{1-2p}\over\Gamma\left(1-p\over2\right)\Gamma\left(p\over2\right)}M.
\label{mMrel}
}
Due to the loss of the conformal invariance the local operators can acquire nonzero vacuum expectation values (VEV). The VEVs of the exponential operators
\eq{
G_a=M^{-2a^2}\cG_a=\langle V_a(x)\rangle\equiv\lvac|V_a(x)|\rvac,
\label{Gadef}
}
are known exactly~\cite{Lukyanov:1996jj,Fateev:1997yg}. It is convenient to write down the factor $\cG_a$ in the form
\Align{
\cG_a
&=\e^{-2\gamma_Ea^2}\cos{\pi a\over Q}
\?\\
&\quad\times\prod^\infty_{k=1}{\e^{2a^2\over k}\Gamma^2\left({1\over2}+{pk\over2}\right)
\Gamma^2\left({1\over2}+{(1-p)k\over2}\right)
\over\Gamma\left({1\over2}-{a\over Q}+{pk\over2}\right)
\Gamma\left({1\over2}+{a\over Q}+{pk\over2}\right)
\Gamma\left({1\over2}-{a\over Q}+{(1-p)k\over2}\right)
\Gamma\left({1\over2}+{a\over Q}+{(1-p)k\over2}\right)},
\quad
0<p<1,
\label{Ga-shG}
}
with $\gamma_E$ being the Euler--Mascheroni constant. This form makes it apparent that the VEV as a function of $a$ possesses no poles, while its zeros, being simple for generic values of~$b$, are located at the points
\eq{
\text{Zeros: }a=\pm a_{mn},\ m\ge n\ge2,\ n\in2\Z\text{ or }n>m\ge2,\ m\in2\Z.
\label{shG-Ga0}
}
The VEV as an analytic function satisfies the equations~\cite{Fateev:1997yg}:
\eq{
G_a=R_aG_{Q-a},
\qquad
G_a=G_{-a}.
\label{Gaprops}
}

Below we shall need the notion and elementary properties of form factors of a local operator. Consider a model with one particle, like the sinh-Gordon model, on the plane. Let  $|\theta_1,\ldots,\theta_N\rangle$ be the stationary state of $N$ particles with the rapidities $\theta_1>\cdots>\theta_N$, so that $|\rvac=|\varnothing\rangle$. Then the form factors of a local (or quasilocal) operator $\cO(x)$ are, by definition, the matrix elements
$$
\langle\theta'_1,\ldots,\theta'_N|\cO(0)|\theta_1,\ldots,\theta_M\rangle.
$$
In principle, for an integrable model the form factors can be found exactly, up to an unknown common factor, by solving a system of linear functional equations~\cite{Smirnov:1992vz}. The words `in principle' mean that there is no straightforward way to identify a solution to the equations with some particular local operator in the Lagrangian approach.

For the exponential operators the form factors were found in~\cite{Koubek:1993ke,Lukyanov:1997bp}. The common factor is fixed by the VEV, so that
\eq{
\langle\theta'_1,\ldots,\theta'_N|V_a(x)|\theta_1,\ldots,\theta_M\rangle
=G_af_a(\theta_1,\ldots,\theta_M,\theta'_N+\i\pi,\ldots,\theta'_1+\i\pi),
\label{fadef}
}
where $f_a$ are analytic functions of the rapidities and $f_a(\varnothing)=1$. To distinguish these functions from the full form factors we shall call them \emph{relative} form factors. The relative form factors are dimensionless and, hence, $\mu$-independent. It is important to note that in the case of the sinh-Gordon model the relative form factors are finite for all values of~$a$, so that the exponential operator $V_a$ just vanishes at zeros of the function~$G_a$:
\eq{
V_{2+2l,k+2+2l}(x)=V_{k+2+2l,2+2l}(x)=V_{-2l,-k-2l}(x)=V_{-k-2l,-2l}(x)=0,
\qquad
k,l=0,1,2,\ldots
\label{Vmn=0}
}
Besides, in contrast to the VEVs, the relative form factors depend on $b$ meromorphically on the whole complex plane.

Later we shall also need the notion of relative form factors extended to descendant operators. For any level $(L,\bar L)$ descendant (never mind, in the Heisenberg or Virasoro sense) $V^\cD_a(x)$ of the exponential operator $V_a(x)$ the form factors can be written in the form
\eq{
\langle\theta'_1,\ldots,\theta'_N|V^\cD_a(x)|\theta_1,\ldots,\theta_M\rangle
=G^\cD_af^\cD_a(\theta_1,\ldots,\theta_M,\theta'_N+\i\pi,\ldots,\theta'_1+\i\pi).
\label{fOadef}
}
Here $G^\cD_a$ is any normalization function proportional to $({\rm mass})^{2\Delta^0_a+L+\bar L}$. It makes the relative form factors dimensionless. We shall try to choose the form of the normalization factor in such a way that it took over as much nonanalyticities in $b$ as possible. Below we shall make use of the conjecture that the relative form factors are analytic in $b$ providing physically relevant answers for both real and imaginary values of~$b$. One of the convenient choices of the normalization factor is $M^{(L+\bar L)}G_a$, but we shall also consider other possibilities.

The sinh-Gordon model can be considered as a perturbation of the Liouville model in two ways. First, we may consider the term with $\e^{b\varphi}$ as a part of the Liouville action, while considering the term with $\e^{-b\varphi}$ as a perturbation. Second, we may vice versa consider $\e^{-b\varphi}$ as a part of the Liouville theory and $\e^{b\varphi}$ as a perturbation. The two perturbation theories are related by the $\varphi\to-\varphi$ symmetry of the model. The $\varphi\to-\varphi$ invariance imposes some property on the form factors of the operators. For the exponential operators it looks like~\cite{Koubek:1993ke}:
\eq{
\langle\theta'_1,\ldots,\theta'_N|V_{-a}(x)|\theta_1,\ldots,\theta_M\rangle
=(-)^{M+N}\langle\theta'_1,\ldots,\theta'_N|V_a(x)|\theta_1,\ldots,\theta_M\rangle.
\label{a0refl}
}
The second of the relations (\ref{Gaprops}) is a particular case of this property. In other words, the particle creation/annihilation operators~\cite{Lashkevich:1994qf} are odd in the field $\varphi$.

The first Liouville perturbation theory imposes the identification of the operators~(\ref{RR}) on the sinh-Gordon theory. On the other hand, the second perturbation theory imposes another identification, which is obtained from~(\ref{RR}) by the substitution $a\to-a$, $Q\to-Q$. Hence, the exponential operators of the sinh-Gordon model satisfy the identities
\subeq{
\label{RR-shG}
\Align{
V_a(x)
&=R_aV_{Q-a}(x),
\label{RR-shG-a}
\\
V_a(x)
&=R_{-a}V_{-Q-a}(x).
\label{RR-shG-b}
}}
By using the properties of the VEVs (\ref{Gaprops}) we can rewrite these identities in the form
\eq{
G_{a'}V_a(x)=G_aV_{a'}(x),
\text{ if }a'-a\in2\Z Q\text{ or }a+a'\in(2\Z+1)Q.
\label{RR-VaVa'}
}
Note that the reflection property for the descendants (\ref{RR-desc}) imposes relations similar to (\ref{RR-shG}) on the descendant operators of the sinh-Gordon model as well.

As it was shown in~\cite{Feigin:2008hs} the form factors obtained in~\cite{Lukyanov:1997bp} for the exponential operators are consistent with these identities:
\eq{
f_a(\theta_1,\ldots,\theta_N)=f_{a'}(\theta_1,\ldots,\theta_N),
\text{ if }a'-a\in2\Z Q\text{ or }a+a'\in(2\Z+1)Q.
\label{RR-shG-ff}
}
This symmetry property affects essentially the interpretation of the analytic function $G_a$. In fact, it can only be interpreted as a vacuum expectation value on the strip
\eq{
-Q/2\le\Re a\le Q/2.
\label{shG-physstrip}
}
On this strip the exponential operators are normalized by the condition of the form
\eq{
\langle V_{a'}(x')V_a(x)\rangle
=R_0^{-1}{\delta_{a+a',Q}+\delta_{a+a',-Q}+R_{\tilde a}\delta_{aa'}\over|x'-x|^{4\tilde a(Q-\tilde a)}}
\quad\text{as $|x'-x|^2\to0$,}
\qquad
\tilde a=a\sign\Re a.
\label{shG-normalization}
}
The first two terms in the numerator are sensible only on the boundary lines $\Re a=\pm Q/2$. The points $a=\pm Q/2=\mp a_{22}$ are exceptional, because $R_{Q/2}=-1$ and, hence, $V_{\pm Q/2}(x)=0$. It is easy to check that the true primary operators at these points are~$V'_{\pm Q/2}(x)$. All other zeros of the VEV lie off the strip, where there is no reasonable normalization condition. We may think the off-strip operators to be just defined by the reflection equations~(\ref{RR-shG}) in terms of the on-strip operators. The zeros may be considered as a result of the unhappy normalization, though there is no way to improve this condition so that zeros would disappear without loss of the analyticity of the~VEV.

Another issue is related to the fact that the reflection function $R_a$ possesses poles at the points
\eq{
\text{Poles of $R_a$: }a=a_{k0},\ a=a_{0k},\ k=1,2,3,\ldots
\label{RaPoles}
}
The poles $a_{k0}$ with $k\le 1+b^2$ and $a_{0k}$ with $k\le1+b^{-2}$ lie on the strip $0\le a\le Q/2$. This means that the short-distance correlation function $\langle V_a(x')V_a(x)\rangle$ for these values of $a$ rises faster than $|x'-x|^{-4a(Q-a)}$ as $|x'-x|^2\to0$. These fields can be normalized by the condition (see Appendix~\ref{App:log})
\eq{
\langle V_a(x')V_a(x)\rangle
=-{2\tilde a\over R_0R'_{Q-\tilde a}}{\log|x'-x|^2\over|x'-x|^{4\tilde a(Q-\tilde a)}}
\quad\text{for $a=\pm a_{k0}$ or $\pm a_{0k}$, $k=1,2,3,\ldots$}
\label{log-shG-normalization}
}
Such fields are very peculiar and deserve a separate study. They correspond to the logarithmic fields in the logarithmic minimal models. The existence of such fields was noticed, in particular, in~\cite{Yamaguchi:2002rt}.

Due to finiteness of the VEV $G_a$ it is safe to say that there is no resonances for the exponential operators themselves. For even values of $m$ it is easy to check that whenever the resonance condition for an exponential field is satisfied, the operator in the right hand side of the resonance equation is a descendant of one of the vanishing operators~(\ref{Vmn=0}). The absence of resonances for odd values of $m$ means that the corresponding perturbation contributions vanish. In the next section we shall see that this fact plays an important role in the sine-Gordon theory.

Let us try to find the resonance of a level $(N,N)$ descendant of the exponential operator $V_a(x)$ with a level $(N',N')$ descendant of the exponential operator $V_{a+(k-l)b}(x)$ in the $(k+l)$th order of the perturbation theory, $k,l=0,1,2,\ldots$, $k+l>0$. It means that $V_{a+(k-l)b}(x)$ must appear in the operator product expansion of the initial descendant operator with $k$ instances of the $\e^{b\varphi}$ operator and $l$ instances of the $\e^{-b\varphi}$ operator. We have the resonance condition in the form
\eq{
\Delta^0_a+N=\Delta^0_{a+(k-l)b}+N'+(k+l)(1+b^2).
\label{shG-resonance}
}
The solution to this equation reads
\eq{
a={1\over2}{k+l+N'-N\over k-l}b^{-1}+{1\over2}\left({k+l\over k-l}-k+l\right)b.
\label{shG-res-solution}
}
Note that this immediately discards the case $k=l$ as a resonance one.

We are not planning to consider all this large selection of possible resonances, but specialize to some subclasses. First of all, let us restrict ourselves to the values $a=a_{mn}$ with positive integers $m,n$ as it was defined in~(\ref{amndef}). In this case $k+l$ and $N-N'$ must be divisible by $k-l$ and we obtain
\subeq{\label{kleq}
\Align{
k+l
&=(l-k)(l-k+n-1),
\label{kleq(k+l)}
\\
N-N'
&=(l-k)(l-k+n-m).
\label{kleq(l-k)}
}}

Consider the resonances inherited from the Liouville theory. These resonances are related to the operators $\Phi_{a|mn}$ as $a\to a_{mn}$. For these operators $N=mn$, which admits a set of solutions to the equations~(\ref{kleq}). The solution $k-l=n$ gives $l=0$, $N'=0$ and corresponds to the contribution of the same field $V_{m,-n}$ as in the Liouville theory. To find other solutions let
$$
l-k=m-s,
$$
which corresponds to the operator $V_{m,n+2m-2s}$. Then we can impose the conditions
\eq{
\Gathered{
N'=s(m+n-s)\ge0,
\\
k={(m-s)(m+n-s-2)\over2}\ge0,
\qquad
l={(m-s)(m+n-s)\over2}\ge0,
\\
k,l\in\Z,
\qquad
k+l>0.
}\label{klms}
}
From these inequalities we immediately obtain
$$
0\le s<m,
\qquad
m-s\in2\Z
\text{ or }
m+n-s\in2\Z.
$$
Besides, from (\ref{shG-Ga0}) we see that $V_{m,n+2m-2s}=0$ for $m\in2\Z$. Since $N'<m(n+2m-2s)$ we conclude that the corresponding descendant vanishes as well and does not contribute the pole. Hence, nonzero contributions may only appear for odd~$m$.

Now recall that the contents of both the Liouville and the sinh-Gordon theories does not change, if we substitute $b\to1/b$ for a fixed value of the parameter~$M$ (or of the mass $m$)~\cite{Fring:1992pt,Zamolodchikov:1995aa}. This means that every nonzero contribution to the resonance residue must be reproduced in the perturbation theory after this substitution. It means that if we have a nonzero contribution from a level $(N',N')$ descendant of $V_{m,n+2m-2s}$, there must be a nonzero contribution from the same operator in the $\ch b^{-1}\varphi$ perturbation theory. But in the latter, the nonzero contributions can come from level $(N',N')$ descendants operators $V^{(s')}_{m+2n-2s',n}$, where
$$
N'=s'(m+n-s'),
\qquad
0\le s'<n,
\qquad
n-s'\in2\Z
\text{ or }
m+n-s'\in2\Z.
$$
To reconcile both perturbation theories, we require each operator $V^{(s)}_{m,n+2m-2s}$ that give a nonzero contribution to coincide (up to a constant factor) to an operator $V^{(s')}_{m+2n-2s',n}$. There are two possibilities. The first one is $s+s'=m+n$, since
$$
a_{m,n+2m-2s}-a_{m+2n-2s',n}=(n-s')b^{-1}+(s-m)b\in2\Z Q,\text{ if $m+n=s+s'$ and $s-m\in2\Z$,}
$$
but this is not the case due to its inconsistency with the conditions $s<m$, $s'<n$.

The second possibility, which is consistent, is $s'=s$. It imposes some restrictions to the values of~$s$. First, $s$ must be less than both $m$ and $n$. Second,
$$
a_{m,n+2m-2s}+a_{m+2n-2s,n}=(1-m-n+s)Q\in(2\Z+1)Q,\text{ if $m+n-s\in2\Z$},
$$
and the operators $V^{(s)}_{m,n+2m-2s}$ and $V^{(s)}_{m+2n-2s,n}$ can only coincide for even $m+n-s$.

Besides, the operator $V^{(s)}_{m,n+2m-2s}$ vanishes for even $m$, while $V^{(s)}_{m+2n-2s,n}$ vanishes for even~$n$. Hence, these operators can only make a nonzero contribution when both $m$ and~$n$ are odd.

Let us summarize the results. There are two types of resonance contributions:
$$
\Aligned{
&{\rm R}_+:\ \Phi_{a_{mn}|mn}\rightsquigarrow V_{m,-n},\ k=n,\ l=0;
\\
&{\rm R}_-^s:\ \Phi_{a_{mn}|mn}\rightsquigarrow V^{(s)}_{m,n+2m-2s}\ \text{with $N',k,l$ from (\ref{klms}) and $0\le s<m,n$,\quad $s\in2\Z$, if $m,n\in2\Z+1$.}
}
$$
Here the sign $\rightsquigarrow$ means `is in resonance with' and $V^{(s)}_{m,n+2m-2s}$ is some level $(N',N')$ descendant of the operator $V_{m,n+2m-2s}$.

The ${\rm R}_+$ resonance is the same as in the Liouville theory and, since it originates from just one term in the perturbation expansion and this term is the same as in the Liouville theory, it is perturbatively exact and the coefficient at the field $V_{m,-n}$ coincides with that obtained in~\cite{Zamolodchikov:2003yb}. The ${\rm R}_-^s$ resonances are specific for the sinh-Gordon model, but they are perturbatively exact as well.

To write all the resonance identities in a compact form it is convenient to define the sets
$$
Z_{mn}=\Cases{\{s\in2\Z\>|\>0\le s<m,n\}&\text{for $m,n\in2\Z+1$,}\\
\varnothing&\text{otherwise.}}
$$
Then the higher equations of motion in the form (\ref{HEM-res}) are modified as follows
\eq{
\Ress{mn}\Phi_{a|mn}(x)=2B_{mn}V_{m,-n}(x)
-\sum_{s\in Z_{mn}}2C^{(s)}_{mn}V^{(s)}_{m,n+2m-2s}(x),
\label{HEM-res-shG}
}
where $C^{(s)}_{mn}=\hmu^{(m-s)(m+n-s-1)}\cC^{(s)}_{mn}$ are some coefficients. The numbers $C^{(s)}_{mn}$ depend on the normalization of the operators $V^{(s)}_{m,n+2m-2s}$. Since the explicit form of these operators is unknown, they are arbitrary except for the case $s=0$, where we may identify $V^{(0)}_{m,n+2m}(x)=V_{m,n+2m}(x)$.

For even $m$ or $n$ the resonance equation retains the form of the higher equations of motion of the Liouville theory:
\eq{
\Ress{mn}\Phi_{a|mn}(x)=2B_{mn}V_{m,-n}(x),
\text{ if $m\in2\Z$ or $n\in2\Z$.}
\label{HEM-res-shG-mneven}
}
For odd $m$ and $n$ the cases $m=1$ and $n=1$ are the simplest ones. We have
\eq{
\Ress{mn}\Phi_{a|mn}(x)=2B_{mn}V_{m,-n}(x)-2C_{mn}V_{m,n+2m}(x),
\text{ if $m,n\in2\Z+1$ and $m=1$ or $n=1$,}
\label{HEM-res-shG-mn1}
}
where the coefficients $C_{mn}=C^{(0)}_{mn}$ can be found explicitly. In principle, we could find them by taking appropriate multiple integrals coming from two-point correlation functions, but it is possible to circumvent explicit integrating. We can make use of the fact that in this case the descendant level $mn$ of the operator $\Phi_{a|mn}$ is odd. On the other hand, for a generic value of $a$ the Fock module coincides with the Verma module of the Virasoro algebra. Accordding to the widely accepted conjecture (see e.g.~\cite{Smirnov:1998-DocMath183}), it is spanned by the vectors generated by the commutative integrals of motion $\cI_{-2k+1}$ of odd spins $2k-1$ and by the Virasoro elements~$\cL_{-2k}$ of even spins~$2k$. The action of the operator $\cI_{-2k+1}$ on an operator $\cO(x)$ is just the commutator with the integral of motion $I_{-2k+1}$ in the space of states: $\cI_{-2k+1}\cO(x)=[\cO(x),I_{-2k+1}]$. Since the vacuum is annihilated by all commuting integrals of motion, we have $\langle\cI_{-2k+1}\cO(x)\rangle=\langle[\cO(x),I_{-2k+1}]\rangle=0$. As far as every Virasoro descendant of an odd level can be represented as a linear combination of commutators of integrals of motion with other operators, its VEV is zero. Hence, $\langle\,\Res_{\,a=a_{mn}}\Phi_{a|mn}(x)\rangle=0$, whence it follows that the VEV of the r.~h.~s.\ of (\ref{HEM-res-shG}) must be equal to zero. The last condition is satisfied, if
\eq{
C_{mn}=B_{mn}{G_{m,-n}\over G_{m,n+2m}}
=B_{mn}{G_{m,-n}\over G_{m,n+2m}},\
\text{if $m,n\in2\Z+1$ and $m=1$ or $n=1$.}
\label{Cmn-oo}
}
Since $a_{m,n+2m}+a_{m,-n}=(1-m)Q\in2\Z Q$ we obtain from (\ref{RR-VaVa'}) that
$$
V_{m,n+2m}(x)={G_{m,n+2m}\over G_{m,-n}}V_{-a_{m,-n}},
\text{ if }m\in2\Z+1.
$$
Hence, we can rewrite the equation (\ref{HEM-res-shG-mn1}) in the form
\eq{
\Ress{mn}\Phi_{a|mn}(x)
=2B_{mn}(V_{a_{m,-n}}(x)-V_{-a_{m,-n}}(x)),
\ m,n\in2\Z+1,\ m=1\text{ or }n=1,
\label{HEM-res-shG-oo}
}
which nicely generalizes the standard equation of motion of the sinh-Gordon model corresponding to $m=n=1$. Some checks of (\ref{HEM-res-shG-mneven}), (\ref{HEM-res-shG-oo}) within the form factor approach are given in Appendix~\ref{App:FF-descendants}.

Now we are ready to define the operators $\Psi_{mn}(x)$ in analogy with those in the Liouville field theory. It is convenient to write the dimensional parameter $\lambda$, defined in (\ref{res-expansion}), as $\kappa M$, where $\kappa$ is an arbitrary dimensionless constant. Then we get the expansion
\Multline{
\Phi_{a|mn}(x)
={2\over a-a_{mn}}\Biggl(B_{mn}(\kappa_+M)^{4nb(a-a_{mn})}V_{a+nb}(x)
\\
-\sum_{s\in Z_{mn}}C^{(s)}_{mn}
(\kappa^{(s)}_-M)^{-4(m-s)b(a-a_{mn})}V^{(s)}_{a-(m-s)b}(x))\Biggr)
\\
+\Psi_{mn}(\underline\kappa;x)+O(a-a_{mn}),
\label{shG-Psi}
}
where $\underline\kappa$ denotes a vector made of all $\kappa$ variables; here $\underline\kappa=(\kappa_+,\kappa_-^{(0)},\kappa_-^{(2)},\ldots,\kappa_-^{(\min(m-1,n-1))})$. Let $\underline1=(1,1,\ldots,1)$. We shall denote $\Psi_{mn}(\underline1;x)=\Psi_{mn}(x)$. Now we want to rewrite this expansion in terms of the relative form factors. Define the relative form factors $f^\Phi_{a|mn}$, $f^\Psi_{mn}$ of the operators $\Phi_{a|mn}$, $\Psi_{mn}$ by the equations
\eq{
\Gathered{
\lvac|\Phi_{a|mn}(0)|\vec\theta\rangle
=\hmu^{n+2n(a-a_{mn})/Q}G_{a+nb}f^\Phi_{a|mn}(\vec\theta),
\\
\lvac|\Psi_{mn}(\underline\kappa;0)|\vec\theta\rangle
=\hmu^nG_{m,-n}f^{\Psi(\underline\kappa)}_{mn}(\vec\theta),
}\label{PhiPsi-fdef}
}
where $\vec\theta$ is a short notation for the sequence $\theta_1,\ldots,\theta_N$ of rapidities of $N$ particles. Besides, define the relative form factors of the operators $V^{(s)}_{a-b(m-s)}$ as
\eq{
\lvac|V^{(s)}_{a-(m-s)b}(0)|\vec\theta\rangle
=M^{2s(m+n-s)}G_{a-(m-s)b}f^{(s)}_{a-b(m-s)}(\vec\theta).
\label{Va-(m-s)b-fdef}
}

The prefactors are chosen in such a way that the relative form factors were dimensionless and finite. By substituting them into (\ref{shG-Psi}) we obtain
\Multline{
f^\Phi_{a|mn}(\vec\theta)
={2\over a-a_{mn}}\left(\cB_{mn}f_{a+nb}(\vec\theta)
-\sum_{s\in Z_{mn}}\cC^{(s)}_{mn}
{\cG_{a-(m-s)b}\over\cG_{a+nb}}f^{(s)}_{a-(m-s)b}(\vec\theta)\right)
\\
+f^\Psi_{mn}(\vec\theta)+O(a-a_{mn}).
\label{shG-Psi-ff}
}
For even $m$ or $n$ the expression is simple:
\eq{
f^\Phi_{a|mn}(\vec\theta)
={2\cB_{mn}f_{a+nb}(\vec\theta)\over a-a_{mn}}
+f^\Psi_{mn}(\vec\theta)+O(a-a_{mn}),
\label{shG-Psi-ff-mneven}
}
while for odd $m,n$ and $m=1$ or $n=1$ it is not much more complicated:
\eq{
f^\Phi_{a|mn}(\vec\theta)
={2\cB_{mn}(f_{a+nb}(\vec\theta)-\cG_{a+nb}^{-1}\cG_{a-mb}f_{a-mb}(\vec\theta))
\over a-a_{mn}}
+f^\Psi_{mn}(\vec\theta)+O(a-a_{mn}).
\label{shG-Psi-ff-mn1}
}
The $\underline\kappa$-dependence is given by the rule
\Multline{
f^{\Psi(\underline\kappa)}_{mn}(\vec\theta)
=f^\Psi_{mn}(\vec\theta)
-8nb\cB_{mn}f_{m,-n}(\vec\theta)\log\kappa_+
\\
-\sum_{s\in Z_{mn}}8b(m-s)\cC^{(s)}_{mn}\cG_{a+nb}^{-1}\cG_{a-(m-s)b}
f_{m,n+2m-2s}(\vec\theta)\log\kappa^{(s)}_-.
\label{shG-Psi-kappa-ff}
}
In the operator language the transformation rule reads
\Multline{
\Psi_{mn}(\underline\kappa;x)=\Psi_{mn}(x)-8nbB_{mn}V_{m,-n}(x)\log\kappa_+
\\*
-\sum_{s\in Z_{mn}}8b(m-s)C^{(s)}_{mn}V_{m,2m+n-2s}(x))\log\kappa^{(s)}_-.
\label{shG-Psi-kappa}
}
This transformation rule incorporates the logarithmic ambiguity of the definition of the renormalized operator. From the form factor point of view the operator $\Psi(x)$ looks most natural, but it is necessary to stress that all operators $\Psi(\underline\kappa;x)$ are, in principle, equitable. Some particular values of $\underline\kappa$ can be distinguished by some particularly nice properties. For example, if $m$ or $n$ is even, the choice $\log\kappa_+=f^\Psi_{mn}(\varnothing)/8nb\cB_{mn}$ makes the VEV of the renormalized operator vanish.

\section{Resonances in the sine-Gordon model}
\label{Sec:Res-sG}

Formally, the sine-Gordon model with the action
\eq{
S_{sG}[\varphi]
=\int d^2x\,\left({(\d_\mu\varphi)^2\over16\pi}+2(-\mu)\cos\beta\varphi\right),
\qquad
0<\beta^2\le1,\ \mu<0,
\label{sG-action}
}
can be obtained from that of the sinh-Gordon theory by the analytic continuation to imaginary $b=-\i\beta$ for fixed~$\hmu$. In terms of the parameter $p$ this theory corresponds to
$$
p<0.
$$
Surely, we want the mass $m$ to be a real positive number. Then according to (\ref{mMrel}) the parameter $M$ must be a complex number with the argument $\i\pi(p-1)$. This is consistent but inconvenient. Thus we make the substitution
$$
M\to\e^{\i\pi(p-1)}M,
$$
so that the new parameter $M$ be real. We shall mean by $M$ this real number in the rest of this section.

Some quantities can be indeed obtained by the analytic continuation. Nevertheless, there are some essential differences between the two models. Physically, these differences are the result of different particle spectra. The sinh-Gordon model contains the only particle, which can be identified with the boson~$\varphi$. The sine-Gordon model for any admissible values of $\beta$ contains a pair of topological solitons --- kink and antikink, which can be considered as elementary particles. For $\beta^2<1/2$ they form a series of bound states --- breathers, the lightest of which, the first breather, can be identified with the `continuation' of the sinh-Gordon particle.

We retain the notation $V_a(x)$ for the exponential operators, though for physically admissible operators the parameter $a$ is purely imaginary, $a=\i\alpha$. The degenerate points $a_{mn}$ correspond to
$$
\alpha_{mn}={1-m\over2}\beta^{-1}-{1-n\over2}\beta.
$$
The form factors of the exponential operators can be again factorized into the VEV and the relative form factors:
\eq{
\lvac|V_a(0)|\theta_1\ve_1,\ldots,\theta_N\ve_N\rangle
=G_af_a(\theta_1,\ldots,\theta_N)_{\ve_1\ldots\ve_N},
\qquad
f_a(\varnothing)=1,
}
where the variables $\ve_i=+,-,1,2,\ldots,\lceil|p|^{-1}-1\rceil$ denote the sorts of particles, $+$ for the kink, $-$ for the antikink, $1,2,\ldots$ for the breathers in order of increasing mass. The relative form factors that only contain the first breather are obtained from those of the sinh-Gordon model by the analytic continuation in the parameter~$b$:
\eq{
f_a(\theta_1,\ldots,\theta_N)_{1\ldots1}
=\left.f_a(\theta_1,\ldots,\theta)\right|_{b=-\i\beta}.
\label{ffident}
}
Moreover, we conjecture that the first breather relative form factors for any local operator can be analytically continued from the sinh-Gordon case.

Nevertheless, since the kink form factors do not satisfy any reflection conditions like (\ref{RR-shG-ff}), the equations (\ref{RR-shG}) do not hold and the operators $V_a$, $V_{Q-a}$ and $V_{-Q-a}$ are essentially different.

The vacuum expectation value $G_a=\langle V_a\rangle$ is similarly represented as $M^{-2a^2}\cG_a$, but the factor $\cG_a$ for the sine-Gordon model cannot be obtained by analytic continuation of that in the sinh-Gordon theory. The obstacle is the fact that the set of zeros (\ref{shG-Ga0}) become everywhere dense for negative $p$ in generic position. It was conjectured in~\cite{Fateev:1997yg} that the VEV has the same form in both cases being written in the integral form, but the analytic structure of this integral in both regions is completely different. Rendered to the form of an infinite product of gamma functions the factor $\cG_a$ reads as
\eq{
\cG_a=\e^{-2\gamma_Ea^2}
\prod^\infty_{k=1}\e^{2a^2\over k}{\Gamma^2\left({1\over2}+{(1-p)k\over2}\right)
\Gamma\left({1\over2}-{a\over Q}-{pk\over2}\right)
\Gamma\left({1\over2}+{a\over Q}-{pk\over2}\right)
\over\Gamma^2\left({1\over2}-{pk\over2}\right)
\Gamma\left({1\over2}-{a\over Q}+{(1-p)k\over2}\right)
\Gamma\left({1\over2}+{a\over Q}+{(1-p)k\over2}\right)},
\quad
p<0.
\label{Ga-sG}
}
This function also satisfies the equations (\ref{Gaprops}), but already possesses poles. It is easy to list zeros and poles of this function:
\eq{
\Aligned{
&\text{Zeros:}
&a
&=\pm a_{mn},
\qquad m>n\ge2,\ m\in2\Z+1,\ n\in2\Z;
\\
&\text{Poles:}
&a
&=\pm a_{mn},
\qquad
m>n\ge1,\ m\in2\Z,\ n\in2\Z+1\text{ or }m\ge2,\ n\le0,\ m\in2\Z,\ n\in\Z.
}\label{Ga-sG-zerpol}
}
The interpretation of the zeros is quite different from the case of the sinh-Gordon model. In that case the relative form factors $f_a(\cdots)$ were finite, and a zero of the VEV meant the whole operator vanish. It was the result of an unhappy normalization of the corresponding exponential operator off the `physical' strip~(\ref{shG-physstrip}). In the sine-Gordon model the normalization condition
\eq{
\langle V_{-a}(x')V_a(x)\rangle\simeq{1\over|x'-x|^{4\Delta^0_a}}
\quad\text{as $|x'-x|^2\to0$}
\label{sG-normcond}
}
is not so vulnerable as~(\ref{shG-normalization}): it makes sense for any value of~$a$. It means that no exponential operator can vanish, and, hence, if the VEV of the operator $V_a$ for some particular value of $a$ vanishes, some of its relative form factors tend to infinity at that value, so that the corresponding absolute form factors remain nonzero. The relative form factors that only contain breathers are always finite. Hence, the infinities must appear for the soliton (kink--antikink) relative form factors. Lukyanov's integral representation~\cite{Lukyanov:1997bp} for the soliton relative form factors of the exponential operators is convergent on the strip
\eq{
|\Im a|<\beta^{-1}-\beta/2
\label{ff-convergence}
}
and diverges as $a\to\pm\i(\beta^{-1}-\beta/2)=\pm a_{32}$. The values $a=\pm a_{32}$, which bound the strip, are, indeed, the closest to the real axis zeros of the VEV. Every other zero of the VEV lies off the strip (\ref{ff-convergence}) all the more. This explains how an exponential operator can occur nonzero even if its VEV vanishes.

The existence of poles suggests the presence of resonances for exponential operators. To verify it consider the conditions (\ref{kleq}) for $N=0$. Let $l-k=m-n-s$. Then $k+l=(m-s-1)(m-n-s)$ and, hence,
$$
k={(m-s-2)(m-n-s)\over2},
\qquad
l={(m-s)(m-n-s)\over2},
\qquad
N'=s(m-n-s).
$$
It can provide resonances for $m>n$, $m\ge2$, if either $m-s\in2\Z$ or $m-n-s\in2\Z$ and $0\le s<m,m-n$.%
\footnote{There is also a set of solutions for $m\le0$, but it can be obtained from this set by means of the substitution $m\to2-m$, $n\to2-n$, which corresponds to $a_{mn}\to-a_{mn}$.}
This is possible in both cases of even and odd~$m$. But for odd $m$ in the sinh-Gordon model there are no poles even in spite of satisfied resonance equations. Hence, we are sure that the corresponding perturbation integrals vanish, and there is no poles in the sine-Gordon case as well. Hence, we limit ourselves to the case of even~$m$. We get the resonances
\AlignAt3{
V_{mn}\rightsquigarrow W^{(s)}_{m,2m-n-2s},
&\quad 0\le s<m-1,m-n,\ s\in\Z,
&&\text{ if $m\ge2$, $n<m$, $m\in2\Z$, $n\in2\Z+1$, or}
\?\\
&\quad 0\le s<m-1,m-n,\ s\in2\Z,
&&\text{ if $m\ge2$, $n<m$, $m,n\in2\Z$.}
\label{exprescond}
}
Here again $W^{(s)}_{m,2m-n-2s}$ is an $(N',N')$ descendant of $V_{m,2m-n-2s}$ and $W^{(0)}_{m,2m-n}=V_{m,2m-n}$. We expect resonances not only for odd positive $n$, as it can be seen from (\ref{Ga-sG-zerpol}), but for even positive $n$ as well. This suggests that the finite values of VEVs for even $n$ do not mean that all form factors are finite: the kink form factors must possess poles. In other words, we may describe the poles and zeros of $G_a$ as follows:
\eq{
\Aligned{
&\text{`Physical' zeros:}
&a
&=\pm a_{mn},
\qquad m>n\ge2,\ m\in\Z,\ n\in2\Z;
\\
&\text{`Physical' poles:}
&a
&=\pm a_{mn},
\qquad
m>n,\ m\ge2,\ m\in2\Z,\ n\in\Z.
}\label{Ga-sG-zerpol-mod}
}
We assume that at the `physical' zeros the breather-to-soliton form factor ratios vanish, while at the `physical' poles the correlation functions of the exponential operator possess a pole due to the resonance phenomenon. The VEV of the exponential operator $G_a$ is regular at every point that is a `physical' zero and a `physical' pole at the same time.

The resonance identity reads
\eq{
\Ress{mn}V_a(x)
=\sum_{s\in Y_{mn}}D^{(s)}_{mn}W^{(s)}_{m,2m-n-2s},
\label{res(N=0)}
}
where $D^{(s)}_{mn}=\hmu^{(m-s-1)(m-n-s)}\cD^{(s)}_{mn}$ are some coefficients and
\eq{
Y_{mn}=\Cases{\{s\in\Z\>|\>0\le s<m-1,m-n\},&\text{if $m\in2\Z$, $n\in2\Z+1$, $m\ge2$, $n<m$,}\\
\{s\in2\Z\>|\>0\le s<m-1,m-n\},&\text{if $m,n\in2\Z$, $m\ge2$, $n<m$,}\\
\varnothing&\text{otherwise.}}
\label{Y-def}
}
Due to the relation $a_{mn}+a_{m,2m-n}=(m+1)Q\in(2\Z+1)Q$ the breather relative form factors of the operators $V_{mn}$ and $V_{m,2m-n}$ coincide for even $m$:
\eq{
f_{mn}(\vec\theta)_{\vec\ve}=f_{m,2m-n}(\vec\theta)_{\vec\ve},
\qquad
\ve_i=1,2,\ldots
\label{fmn-eq-br}
}
This identity makes it possible to establish the coefficient at the operator~$V_{m,2m-n}$. In the vicinity of the point $a_{mn}$ the VEV admits the expansion
\eq{
G_a=M^{2(a_{mn}^2-a^2)}\left({G^{(-1)}_{mn}\over a-a_{mn}}+G^{(0)}_{mn}+O(a-a_{mn})\right),
\qquad
G^{(\alpha)}_{mn}=M^{2a_{mn}^2}\cG^{(\alpha)}_{mn}.
\label{Gaexpansion}
}
From (\ref{fmn-eq-br}) and (\ref{Gaexpansion}) we get the proportionality coefficient
\eq{
D_{mn}\equiv D^{(0)}_{mn}={G^{(-1)}_{mn}\over G_{m,2m-n}}.
\label{VVeq}
}

Besides, the identity (\ref{fmn-eq-br}) implies that the total contribution of the operators $W^{(s)}_{m,2m-n-2s}$ with $s\ne0$ into the breather form factor is zero. Of course, their contribution to soliton form factors can be nonzero, so that the identity (\ref{fmn-eq-br}) is unlikely to hold for solitons.

To define the regular part consider the expansion
\eq{
V_a(x)
=\sum_{s\in Y_{mn}}
{(\nu^{(s)}M)^{-4(m-n-s)b(a-a_{mn})}D^{(s)}_{mn}W^{(s)}_{a-(m-n-s)b}(x)\over a-a_{mn}}
+\tilde V_{mn}(\underline\nu;x)+O(a-a_{mn}).
\label{tildeVdef}
}
Here $\underline\nu=(\nu^{(s)})_{s\in Y_{mn}}$ are free parameters.

To establish the form factor relations let us define the relative form factors $g^{(s)}$ of the operators $W^{(s)}$ as
\eq{
\lvac|W^{(s)}_{a-(m-n-s)b}(0)|\overrightarrow{\theta\ve}\rangle
=M^{2s(m-n-s)-4sb\left(a-{m-n-2s\over2}b\right)}
G_{a-(m-n)b}\,g^{(s)}_{a-(m-n-s)b}(\vec\theta)_{\vec\ve}.
\label{Wfdef}
}
Substituting it into (\ref{res(N=0)}), we obtain
\eq{
\cG^{(-1)}_{mn}f_{mn}(\vec\theta)_{\vec\ve}
=\cG_{m,2m-n}\sum_{s\in Y_{mn}}\cD^{(s)}_{mn}g^{(s)}_{m,2m-n-2s}(\vec\theta)_{\vec\ve}.
\label{res(N=0)ff}
}
Then from the regular part of (\ref{tildeVdef}) we obtain
\Multline{
\lvac|\tilde V_{mn}(\underline\nu;0)|\overrightarrow{\theta\ve}\rangle
=G^{(0)}_{mn}f_{mn}(\vec\theta)_{\vec\ve}
+G^{(-1)}_{mn}f'_{mn}(\vec\theta)_{\vec\ve}
\\
-M^{-2a_{mn}^2}\sum_{s\in Y_{mn}}\cD_{mn}^{(s)}
\Bigl(\cG_{m,2m-n}g^{(s)\prime}_{m,2m-n-2s}(\vec\theta)_{\vec\ve}
\\
+\left(\cG'_{m,2m-n}-4\cG_{m,2m-n}(m-n-s)b\log\nu^{(s)}\right)
g^{(s)}_{m,2m-n-2s}(\vec\theta)_{\vec\ve}\Bigr),
\label{tildeV-ff}
}
where primes mean derivatives in $a$: $f'_{mn}(\cdots)=\left.{d\over da}f_a(\cdots)\right|_{a=a_{mn}}$ etc. The particular choice
$$
\log\nu^{(s)}=\log\nu_*^{(s)}\equiv
{1\over4(n-m-s)b}\left({\cG'_{m,2m-n}\over\cG_{m,2m-n}}-{\cG^{(0)}_{mn}\over\cG^{(-1)}_{mn}}\right)
$$
is especially nice, since the first term is cancelled by the terms proportional to $g^{(s)}_{m,2m-n-2s}$:
$$
\lvac|\tilde V_{mn}(\underline{\nu_*};0)|\overrightarrow{\theta\ve}\rangle
=G^{(-1)}_{mn}f'_{mn}(\vec\theta)_{\vec\ve}
-M^{-2a_{mn}^2}\cG_{m,2m-n}
\sum_{s\in Y_{mn}}\cD_{mn}^{(s)}g^{(s)\prime}_{m,2m-n-2s}(\vec\theta)_{\vec\ve}.
$$
Below we shall need another choice: $\underline\nu=\underline1$. In the breather sector, due to (\ref{fmn-eq-br}), (\ref{res(N=0)ff}) we have for $\tilde V_{mn}(x)=\tilde V_{mn}(\underline1;x)$:
\eq{
\lvac|\tilde V_{mn}(0)|\overrightarrow{\theta\ve}\rangle
=\tilde G^{(0)}_{mn}f_{mn}(\vec\theta)_{\vec\ve}
+2G^{(-1)}_{mn}f'_{mn}(\vec\theta)_{\vec\ve},
\qquad
\tilde G^{(0)}_{mn}=G^{(0)}_{mn}-G^{(-1)}_{mn}\cG^{-1}_{m,2m-n}\cG'_{m,2m-n}.
\label{tildeV-ff1-br}
}

Now consider the $mn$ level resonances. There are two cases: the case of odd $m$ and that of even~$m$. For odd $m$ the VEV of the operator $V_{m,-n}$ is finite, and the resonances are very similar to those in the sinh-Gordon case. It is easy to see that the only difference from~(\ref{shG-Psi}) is that the sum in the parenthesis must be taken over the set
\eq{
Z'_{mn}=\Cases{\{s\in\Z|0\le s<m\},
&\text{if $m,n\in2\Z+1$,}\\
\{s\in2\Z+1|1\le s<m\},
&\text{if $m\in2\Z+1$, $n\in2\Z$.}
}
\label{Z'-modd-def}
}

The case of even $m$ differs greatly. On one hand, the operators $V_{m,n+2m-2s}$ do not vanish in this case, so that we have to admit an expansion like (\ref{shG-Psi}), but with summation over the set
\eq{
Z'_{mn}=\Cases{\{s\in\Z|0\le s<m\},
&\text{if $m\in2\Z$, $n\in2\Z+1$,}\\
\{s\in2\Z|0\le s<m\},
&\text{if $m,n\in2\Z$.}
}
\label{Z'-meven-def}
}
On the other hand, the first term in (\ref{shG-Psi}) contains the operator $V_{a+nb}$, which itself has a pole at $a=a_{mn}$ according to~(\ref{res(N=0)}),~(\ref{Y-def}), since $a_{mn}+nb=a_{m,-n}$. Besides, this expansion must be consistent with the form factor identity~(\ref{shG-Psi-ff-mneven}) in the breather sector.

First of all, notice that the ratios $G_a/G_{a+nb}$ are meromorphic functions of $b$ on the complex plane, whose explicit form can be extracted from (\ref{Ga-shG}) for real values of~$b$ and from (\ref{Ga-sG}) for imaginary ones. Besides, for generic values of $b$ they do not possess poles at $a=a_{mn}$ for any $m>0$, if $n>0$. Hence, the numbers
\eq{
K_{mn}=\left.G_a\over G_{a+nb}\right|_{a\to a_{mn}},
\qquad
m,n>0,
\label{Kmndef}
}
are well defined:
\Multline{
K_{mn}=(Mb^{2(1-p)})^{2(Qb-m)n}
\\
\times
\prod^n_{k=1}{\Gamma(1-2ba_{m,n+1-2k})\Gamma(1+Q^{-1}a_{m,n+1-2k})\Gamma({1\over2}-Q^{-1}a_{m,n+1-2k})
\over\Gamma(1+2ba_{m,n+1-2k})\Gamma(1-Q^{-1}a_{m,n+1-2k})\Gamma({1\over2}+Q^{-1}a_{m,n+1-2k})}.
\label{Kmnexplicit}
}
Evidently,
\eq{
K_{mn}=0,
\text{ if }\ m\ge n\ge2,\ n\in2\Z\text{ or }n>m\ge2,\ m\in2\Z.
\label{Kmnzeros}
}
As a consequence, the definition of $\Phi_{a|mn}$ in terms of the form factors is reliable in the sine-Gordon model as well. Thus we may continue the identity (\ref{shG-Psi-ff-mneven}) analytically to the imaginary values of~$b$ and multiply it by the quantity $\hmu^{n+2n(a-a_{mn})/Q}G_{a+nb}$ for the sine-Gordon model, which possesses a pole at $a=a_{mn}$. We obtain in the breather sector ($\ve_i=1,2,\ldots$)
\Align{
\lvac|\Phi_{a|mn}(0)|\overrightarrow{\theta\ve}\rangle
&=\hmu^{n+2n(a-a_{mn})/Q}\left({2\cB_{mn}G_{a+nb}f_{a+nb}(\vec\theta)_{\vec\ve}\over a-a_{mn}}
+G_{a+nb}f^\Psi_{mn}(\vec\theta)_{\vec\ve}\right)+O(1)
\?\\
&={2B_{mn}M^{-4mb(a-a_{mn})}G_{m,2m+n}^{-1}G^{(-1)}_{m,-n}G_{a-mb}f_{a-mb}(\vec\theta)_{\vec\ve}
\over(a-a_{mn})^2}
\?\\*
&\quad+{2B_{mn}(2G^{(-1)}_{m,-n}f'_{m,-n}(\vec\theta)_{\vec\ve}
+\tilde G^{(0)}_{m,-n}f_{m,-n}(\vec\theta)_{\vec\ve})
+\hmu^nG^{(-1)}_{m,-n}f^\Psi_{mn}(\vec\theta)_{\vec\ve}\over a-a_{mn}}+O(1)
}
as $a\to a_{mn}$. By comparing with (\ref{tildeV-ff1-br}), after generalization to the soliton sector, we may conjecture
\Align{
\Phi_{a|mn}(x)
&={2B_{mn}M^{4nb(a-a_{mn})}\over(a-a_{mn})^2}
\sum_{s\in Y_{m,-n}}
(\nu^{(s)}M)^{-4(m+n-s)b(a-a_{mn})}D^{(s)}_{m,-n}W^{(s)}_{a-(m+n-s)b}(x)
\?\\*
&\quad+{2B_{mn}\tilde V_{m,-n}(\underline\nu;x)+\Psi_{mn}(x)\over a-a_{mn}}+O(1).
\label{sG-Phi-meven}
}
The finite operator $\Psi$ defined by this expansion possesses in the breather sector the form factors
\eq{
\lvac|\Psi_{mn}|\overrightarrow{\theta\ve}\rangle
=\hmu^nG^{(-1)}_{m,-n}f^\Psi_{mn}(\vec\theta)_{\vec\ve}.
\label{sG-Psi-meven-def}
}
Notice that the operator $\Psi_{mn}$ appears as a finite part of the expansion of $(a-a_{mn})\Phi_{a|mn}$ rather than of $\Phi_{a|mn}$ itself. It means an important circumstance that the operator $\Phi_{a|mn}(x)$ needs a multiplicative renormalization beside an additive one.

\section{Conclusion}
\label{Sec:Conclusion}

We have used some consistency conditions to establish the form of the resonance identities and to define renormalized operators in the Liouville, sinh- and sine-Gordon theories. We found that in the case of the sinh-Gordon model the resonance identities for some operators remain the same as in the Liouville model, while for the other operators the resonance identities are modified getting a finite number of extra terms. These extra terms are perturbatively exact in the sense that for any given correlation function each of them appears in a single term in the perturbation expansion. We have seen that the sine-Gordon model possesses a much more complicated structure of resonances. The resonance identities are expected to be exact operator identities that may help us to establish the correspondence between the bootstrap form factors and local operators of the Lagrangian field theory.

Strictly speaking, most of the expressions given here are conjectures. Their consistency to the form factor identities in some simple cases is encouraging, but we need a more rigorous derivation. An explicit computation of integrals in the perturbation theory would provide a rigorous check for these conjectures.

\section*{Acknowledgments}

I am grateful to A.~Belavin, Vl.~Dotsenko, A.~Litvinov, F.~Smirnov I.~Tipunin, Ya.~Pugai, and C.~Rim for discussions. The work was supported, in part, by the Russian Foundation for Basic Research under the grants 09--02--91064 and 09--02--93106 and by the Program for Support of Leading Scientific Schools under the grant 6501.2010.2. The initial idea of the work appeared during my stay at LPTHE in the fall of 2010 supported by the LIA Physique th\'eorique et mati\`ere condens\'ee (aka ENS--Landau program).

\Appendix

\section{Proof of the normalization condition~(\ref{log-shG-normalization})}
\label{App:log}

Here we derive the normalization condition at the poles of the reflection function~$R_a$ in the framework of the Liouville field theory, where we do not have to restrict the equations to the strip~(\ref{shG-physstrip}).

Let us start with a general remark. The pair correlation functions $\langle V_a(x)V_{Q-a}(0)\rangle$, $\langle V_a(x)V_a(0)\rangle$ correspond to the points where the structure constants possess a pole~\cite{Zamolodchikov:1995aa}. It means that the two-point correlation functions in the Liouville theory formally diverge. Nevertheless, there is a way to define finite two-point functions as residues of some three-point functions. First define the functions
\eq{
\Aligned{
\langle V_\delta(\infty)V_a(x)V_{Q-a-\delta}(0)\rangle
&=\lim_{x_\infty\to\infty}R_0^{-1}|x_\infty|^{2\Delta_\delta}
\Res_{\delta_\infty=\delta}\langle V_{\delta_\infty}(x_\infty)V_a(x)V_{Q-a-\delta}(0)\rangle,
\\
&
=
R_0^{-1}|x|^{2\Delta_\delta-2\Delta_a-2\Delta_{a+\delta}},
\\
\langle V_\delta(\infty)V_a(x)V_{a+\delta}(0)\rangle
&=R_{a+\delta}\langle V_\delta(\infty)V_a(x)V_{Q-a-\delta}(0)\rangle.
}\label{VVV-def}
}
Then the two-point functions are obtained in the limit $\delta\to0$:
\eq{
\Aligned{
\langle V_a(x)V_{Q-a}(0)\rangle
&=\langle V_\delta(\infty)V_a(x)V_{Q-a-\delta}(0)\rangle|_{\delta\to0},
\\
\langle V_a(x)V_a(0)\rangle
&=\langle V_\delta(\infty)V_a(x)V_{a+\delta}(0)\rangle\rangle|_{\delta\to0}.
}\label{VV-def}
}
It explains why we introduced the factor $R_0^{-1}$ into the definition of the residue correlation functions: it is necessary to conform the natural identity $\langle V_0(x)V_0(0)\rangle=1$.

Now turn to the special values of~$a$ we are interested in. Let $a_*$ be any of the points $a_{k0}$ or $a_{0l}$, which are poles of the reflection function. Since $R_{Q-a}=R_a^{-1}$, it has a simple zero at~$a_*$:
$$
R_{Q-a}=-(a-a_*)R'_{Q-a_*}+O((a-a_*)^2)
\quad\text{as $a\to a_*$,}
$$
where $R'_a$ is the derivative of~$R_a$. Now consider the product
$$
\left.R'_{Q-a_*-\delta}V_\delta(\infty)V_{a_*+\delta}(0)\right|_{\delta\to0}
=-\left.{1\over\delta}R_{Q-a_*-\delta}V_\delta(\infty)V_{a_*+\delta}(0)\right|_{\delta\to0}
=-\left.{1\over\delta}V_\delta(\infty)V_{Q-a_*-\delta}(0)\right|_{\delta\to0}\>.
$$
Since all the operators $V_{a_*}(x)$ are finite (because they possess finite three-point functions with nearly all exponential fields), the operators $V_{Q-a_*}(x)$ vanish. Hence, we may substitute the limit in the last expression by a derivative:
$$
R'_{Q-a_*}V_0(\infty)V_{a_*}(0)
=-\left.{d\over d\delta}V_\delta(\infty)V_{Q-a_*-\delta}(0)\right|_{\delta=0}\>.
$$
By using the definition (\ref{VV-def}) we obtain
\eq{
\langle V_{a_*}(x)V_{a_*}(0)\rangle
=\left.\langle V_\delta(\infty)V_{a_*}(x)V_{a_*+\delta}(0)\rangle\right|_{\delta\to0}
=\left.-R^{\prime-1}_{Q-a_*}
{d\over d\delta}\langle V_\delta(\infty)V_{a_*}(x)V_{Q-a_*-\delta}(x)\rangle\right|_{\delta=0}\>.
\?}
Taking the explicit form of the three-point function in the r.~h.~s.\ from~(\ref{VVV-def}) we see that
$$
\langle V_{a_*}(x)V_{a_*}(0)\rangle
=-R^{\prime-1}_{Q-a_*}R_0^{-1}
\left.{d\over d\delta}|x|^{2\Delta_\delta-2\Delta_{a_*}-2\Delta_{a_*+\delta}}\right|_{\delta=0}
=-{2a_*\over R'_{Q-a_*}R_0|x|^{2a_*(Q-a_*)}}\log|x|^2,
$$
which leads to the short range asymptotics~(\ref{log-shG-normalization}) in the sinh-Gordon theory.

\section{Form factors of the level \texorpdfstring{$\le3$}{<=3} descendants and resonance identities}
\label{App:FF-descendants}

In \cite{Feigin:2008hs} a construction for breather form factors in the sine-Gordon model was proposed.%
\footnote{A construction for soliton form factors was also proposed, but it needs some refinement due to problems with the convergence of integrals.} First, we briefly review it in the Liouville/sinh-Gordon notation used in the present paper.

Let $\cA$ be a commutative algebra with the generators $c_{-n}$, $n=1,2,\ldots$, and $\bcA$ be its copy with the generators $\bc_{-n}$. Let $\cA^2$ be the associative algebra generated by elements of both algebras with the additional commutative relation
$$
[c_{-m},\bc_{-n}]={m\over2\sin^2{\pi mp\over2}}\delta_{mn}\times\Cases{0,&m\in2\Z+1,\\1,&m\in2\Z.}
$$
To any element $g$ of the algebra $\cA^2$ we associate a sequence of functions $\{P^h(x_1,\ldots,x_k|y_1,\ldots,y_l)\}^\infty_{k,l=0}$ defined according to the following rule:
\Gather{
P^1(X|Y)=1;
\qquad
P^{c_{-n}}(X|Y)=S_n(X)+(-)^{n-1}S_n(Y);
\qquad
P^{\bc_{-n}}(X|Y)=S_{-n}(Y)+(-)^{n-1}S_{-n}(X);
\?\\
P^{k_1g_1+k_2g_2}(X|Y)=k_1P^{g_1}(X|Y)+k_2P^{g_2}(X|Y),
\quad
k_1,k_2\in\C,\ g_1,g_2\in\cA^2;
\?\\
P^{hh'}(X|Y)=P^h(X|Y)P^{h'}(X|Y),
\quad
h,h'\in\cA;
\qquad
P^{\bh\bh'}(X|Y)=P^\bh(X|Y)P^{\bh'}(X|Y),
\quad
\bh,\bh'\in\bcA;
\?\\
P^{\bh'h}(X|Y)=P^h(X|Y)P^{\bh'}(X|Y),
\quad
h\in\cA,\ \bh'\in\bcA,
\?}
where
$$
X=(x_1,\ldots,x_k),\quad Y=(y_1,\ldots,y_l),\quad S_n(X)=\sum^k_{i=1}x_i^n.
$$
It is important that in the last line the element $\bh'$ \emph{precedes} the element $h$ in the product~$\bh'h$. Let $g=h\otimes\bh'\in\cA\otimes\bcA$. Define the element $[g]=h\bh'\in\cA^2$. The linear map $[\cdot]$ puts a `barred' element to the \emph{right} of the `unbarred'. The form factor construction uses the functions $P^{[g]}(X|Y)$, where we first need to push the `barred' elements through the `unbarred' ones to the left and then apply the above rules.

Define the constant $\lambda'$ and the functions $R(\theta)$ and $f(z)$ according to
\Gather{
\lambda'=\left(1\over2\sin{\pi p\over2}\right)^{1/2}\exp\int^{\pi p}_0dt\,{t\over\sin t},
\?\\
R(\theta)
=\exp\left(
-4\int^\infty_0{dt\over t}\,
{\sh{\pi t\over2}\sh{\pi pt\over2}\sh{\pi(1-p)t\over2}
\over\sh^2\pi t}\ch(\pi-\i\theta)t
\right),
\?\\
f(z)={(z+\omega)(z-\omega^{-1})\over z^2-1},
\qquad
\omega=\e^{-\i\pi p}.
\?}
Then the relative form factors of some descendant $V^g_a(x)$ ($g\in\cA\otimes\bcA$) of the operator $V_a(x)$ are given by
$$
f^g_a(\theta_1,\ldots,\theta_N)=\lambda^{\prime\,N}
J^g_{\nu,N}(\e^{\theta_1},\ldots,\e^{\theta_N})\prod^N_{i<j}R(\theta_i-\theta_j),
\qquad
\nu={a\over Q}-{1\over2},
$$
where $J^g_{\nu,N}(X)$ are the symmetric functions given by
$$
J^g_{\nu,N}(X)=\sum_{X=X_-+X_+}\e^{\i\pi\nu(\#X_--\#X_+)}P^{[g]}(X_-|X_+)
\prod_{x\in X_-\atop y\in X_+}f\left(x\over y\right).
$$
Here the sum is taken over all exact partitions of the (multi)set $X$ into pairs of nonoverlapping sets $X_-$ and~$X_+$. In the simplest case $g=1\otimes1$ the functions $f_a(\vec\theta)=f^{1\otimes1}_a(\vec\theta)$ are the relative form factors of the exponential operators~$V_a(x)$.

Now we are ready to describe several particular cases, related to the resonance operators described in the paper. We can introduce some particular elements $h^{(l)}_\nu$, $l=1,2,3$ on each level $l$ of the form
\Align{
h^{(1)}_\nu
&={c_{-1}\over\cos\pi\nu},
\label{h1def}
\\
h^{(2)}_\nu
&={c_{-2}-\i\tg\pi\nu\>c_{-1}^2\over\sin\pi p+\sin2\pi\nu},
\label{h2def}
\\
h^{(3)}_\nu
&={h^{(2)}_\nu c_{-1}-{\i\over3\cos^2\pi\nu}(c_{-3}-c_{-1}^3)
\over\cos2\pi p+\cos2\pi\nu}.
\label{h3def}
}
We use them to define the operators
\eq{
V^{(l)}_a(x)=\tV^{g^{(l)}_a}_a(x),
\qquad
g^{(l)}_a=h^{(l)}_\nu\bar h^{(l)}_{-\nu}.
\label{V(l)adef}
}
It can be easily checked that the operator
$$
V^{(1)}_a(x)=-\left(2Q\over\pi m\right)^2\d_+\d_-V_a(x)
=-\left(2Q\over\pi m\right)^2D_{11}\bar D_{11}V_a(x)
$$
possesses a simple pole at $a=0$ and
$$
\Res_{a=0}V^{(1)}_a=-\left(2Q\over\pi m\right)^2\d_+\d_-\varphi,
$$
where $m$ is the mass of the first breather~(\ref{mMrel}). The corresponding breather form factors of this operator indeed satisfy the identities\cite{Babujian:2002fi} corresponding to the equation of motion
$$
\d_+\d_-\varphi=4\pi\mu b\sh b\varphi\equiv4\pi(-\mu)\beta\sin\beta\varphi,
$$
which is the resonance identity for $\Phi_{a|11}$ from~(\ref{HEM-res-shG-mn1}). It is easy to obtain the answer for the renormalized operator $\Psi_{11}(x)=\d_+\varphi\d_-\varphi(x)$ in terms of relative form factors of exponential operators:
\Multline{
\langle{\rm vac}|\d_+\varphi\d_-\varphi(0)|\overrightarrow{\theta\ve}\rangle
={1\over8}\left(\sum^N_{i=1}m_{\ve_i}\e^{\theta_i}\right)
\left(\sum^N_{i=1}m_{\ve_i}\e^{-\theta_i}\right)f''_0(\vec\theta)_{\vec\ve}
-2\pi\mu bK_{11}^{-1}(f'_b(\vec\theta)_{\vec\ve}-f'_{-b}(\vec\theta)_{\vec\ve})
\\
-2\pi\mu bK_{11}^{-1}\left((L+4b\log\kappa_+)f_b(\vec\theta)_{\vec\ve}
+(L+4b\log\kappa_-)f_{-b}(\vec\theta)_{\vec\ve}\right).
\label{Psi11ff}
}
Here $m_\ve$ is the mass of the $\ve$th particle (so that $m_1=m$), and $L=\cG'_b/\cG_b$. It seems to be reasonable to choose $\kappa_+=\kappa_-$. For this choice the form factors for the states consisting of the first breathers vanish for odd number of particles. Intuitively, this corresponds to the fact that the operator is even with respect to~$\varphi$.

The most convenient choice is $\log\kappa_+=\log\kappa_-=-L/4b$, since the expression in the last line completely vanishes. For this choice the zero- and one-breather form factors vanish, while the two-breather one reads explicitly
$$
\langle{\rm vac}|\d_+\varphi\d_-\varphi(0)|\theta_11,\theta_21\rangle
=\pi^2p(1-p)\lambda'm^2R(\theta_1-\theta_2)\left(
(\e^{(\theta_1-\theta_2)/2}+\e^{(\theta_2-\theta_1)/2})^2-2\cos\pi p\right).
$$
In the limit $b\to0$, $m=\const$ the form factor tends to its free field value $\pi m^2(\e^{\theta_1-\theta_2}+\e^{\theta_2-\theta_1})$.

The following results have been checked up to the 4-particle form factor analytically and up to the 8-particle form factor numerically.

For $V^{(2)}_a(x)$ we have the conjectural identities
$$
\Ress{mn}V^{(2)}_a(x)={QK_{mn}\over2\pi\sin^3\pi p}V_{a_{m,-n}}(x),
\qquad
mn=12,21,
$$
which can be compared with the $\Phi_{a|12}$ and $\Phi_{a|21}$ resonance identities~(\ref{HEM-res-shG-mn1}). This comparison gives
$$
V^{(2)}_a(x)={QK_{mn}\over4\pi B_{mn}\sin^3\pi p}\Phi_{a|mn}(x)+O((a-a_{mn})^0),
\qquad
mn=12,21.
$$
Of course, we can extract the finite parts of the operator $V^{(2)}_a$, but we cannot be sure that they coincide with the corresponding operators $\Psi_{12}$, $\Psi_{21}$. They may contain contributions from $\cL_{-1}^2\bar\cL_{-1}^2V_{mn}$, $\cL_{-1}^2\bar\cL_{-2}V_{mn}$, $\cL_{-2}\bar\cL_{-1}^2V_{mn}$.

For $V^{(3)}_a(x)$ we can also conjecture some identities. It was checked that the relative form factors only possess simple poles at $a=a_{13},a_{31}$. Then we have
$$
\Ress{mn}V^{(3)}_a(x)
={QK_{mn}\over32\pi\sin^5\pi p\cos^2\pi p}(V_{a_{m,-n}}(x)-V_{-a_{m,-n}}(x)),
\qquad
mn=13,31.
$$

\raggedright

\end{document}